\documentclass[prd,aps,showpacs,epsf,floats,onecolumn]{revtex4}%
\usepackage{amssymb}
\usepackage{amsfonts}
\usepackage{amsmath}
\usepackage{graphicx}%
\providecommand{\U}[1]{\protect\rule{.1in}{.1in}}

\begin{document}
\title{\textbf{APPLICATION OF THE MAXIMUM RELATIVE ENTROPY METHOD TO THE PHYSICS OF
FERROMAGNETIC MATERIALS}}
\author{Adom Giffin$^{1}$, Carlo Cafaro$^{2}$, and Sean Alan Ali$^{3}$}
\affiliation{$^{1}$Clarkson University, 13699 Potsdam, New York, USA}
\affiliation{$^{2}$SUNY Polytechnic Institute, 12203 Albany, New York, USA }
\affiliation{$^{3}$Albany College of Pharmacy and Health Sciences, 12208 Albany, New York, USA}

\begin{abstract}
It is known that the Maximum relative Entropy (MrE) method can be used to both
update and approximate probability distributions functions in statistical
inference problems. In this manuscript, we apply the MrE method to infer
magnetic properties of ferromagnetic materials. In addition to comparing our
approach to more traditional methodologies based upon the Ising model and Mean
Field Theory, we also test the effectiveness of the MrE method on
conventionally unexplored ferromagnetic materials with \emph{defects}.

\end{abstract}

\pacs{Entropy (89.70.Cf), Probability Theory (02.50.Cw), Statistical Mechanics of
Model Systems (64.60.De).}
\maketitle

\section{Introduction}

In 1957, Jaynes \cite{Jaynes, Jaynes2} showed that maximizing statistical
mechanic entropy for the purpose of revealing how gas molecules were
distributed was simply the maximizing of the Shannon information entropy
\cite{Shannon1948} with statistical mechanical information. This idea lead to
MaxEnt or his use of the Method of Maximum Entropy for assigning
probabilities. This method has recently evolved to a more general method, the
method of Maximum relative Entropy (MrE) \cite{CatichaGiffin} which has the
advantage of not only assigning probabilities but \emph{updating} them when
new information is given in the form of constraints on the family of allowed
posteriors. One of the drawbacks of the MaxEnt method was the inability to
include data. When data was present, one used Bayesian methods. The methods
were combined in such a way that MaxEnt was used for assigning a prior for
Bayesian methods, as Bayesian methods could not deal with information in the
form of constraints, such as expected values. Previously it has been shown
that one can use the MrE method to reproduce a mean field solution for a
simple fluid \cite{Tseng}. The purpose of this was to illustrate that in
addition to updating probabilities, MrE can also be used for
\emph{approximating} probability distributions as an approximation tool.

In a simple ferromagnetic material (that is, a ferromagnetic material with a
single domain), the electronic spins of the individual atoms are strong enough
to affect one and other, and give rise to the so called exchange interaction
\cite{Reif}. This effect, however, is temperature dependent. When the
temperature is below a certain point (the Curie or critical temperature) the
spins tend to all point in the same direction due to their influence on each
other. This establishes a permanent magnet as the individual atoms produce a
net dipole effect. Above this temperature, the atoms cease to have a
significant effect on each other and the material behaves more like a
paramagnetic substance. Determining this net dipole effect can be difficult.
First, the interactions are due to complicated quantum effects. Second, since
a given material has a very large number of atoms, computing the net dipole
effect can be difficult in two dimensions and completely intractable in three
dimensions. Therefore, approximations such as using an Ising Model
\cite{lieb64, brush67, cipra00} and/or the mean field approximation
\cite{huang, mermin, Schroeder} are made to facilitate computation.

Applications of the MrE (updating) method together with information geometric
methods used to characterize the complexity of dynamical systems described in
terms of probabilistic tools are quite extensive \cite{carlo1, carlo2, carlo3,
carlo4, carlo5, carlo6, carlo7, carlo8, carlo9}. In Ref. \cite{carlo2,
carlo3}, using the MrE method together with differential geometric techniques,
we proposed an information-geometric characterization of chaotic energy level
statistics of a quantum antiferromagnetic Ising spin chain in a tilted
magnetic field. In Ref. \cite{carlo9}, employing the very same aforementioned
techniques, we were able to establish a connection between the behavior of the
information-geometric complexity of a trivariate Gaussian statistical model
and the geometric frustration phenomena that appears in triangular Ising
models \cite{sadoc06}. However, the purpose of our article is to illustrate
the use of the MrE (approximating) method as a tool for attaining
approximations for ferromagnetic materials that lie outside the ability of
traditional methods. In doing so, we further the previous work done and show
the versatility of the method.

The layout of the remaining part of this manuscript is as follows. In Section
II, we briefly outline the essential steps of the MrE method in updating and
approximating probability distributions. In Section III, we describe the
basics of the Ising model and Mean Field Theory as approximate mathematical
descriptions of ferromagnetic materials. In Section IV, we compare
magnetization properties of ferromagnets inferred by means of MrE with those
obtained via the Ising model together with Mean Field Theory. In Section V, we
further test the effectiveness of the MrE methodology by considering
ferromagnetic material in the presence of defects. Our final remarks appear in
Section VI.

\section{The Maximum relative Entropy method}

In this Section, we outline the essential elements of the MrE method as a
technique for updating and/or approximating probability distributions.

\subsection{Updating probability distributions}

The MrE method is a technique for updating probabilities when new information
is provided in the form of a constraint on the family of the allowed
posteriors. The main feature of the MrE method is the possibility of updating
probabilities in the presence of both data and expected value constraints.
This feature was first formally presented in \cite{CatichaGiffin} where, in
particular, it was shown that Bayes updating can be regarded as a special case
of the MrE method.\ A first semi-quantitative analysis of the effective
advantages of this powerful feature of the MrE method appeared in
\cite{giffin-caticha}. Finally, the first fully quantitative investigation of
the advantages of the MrE method was carried out in \cite{giffin} where two
toy problems were solved in detail. Following these lines of investigation, we
present here a novel application of the MrE method to a real-world
ferromagnetic problem.

We use the MrE method to update from a prior to a posterior probability
distribution. Specifically, we want to make inferences on some quantity
$\theta\in\Theta$ given:

\begin{description}
\item[i)] the prior information about $\theta$ (the prior);

\item[ii)] the known relationship between $D\in\mathcal{D}$ and $\theta
\in\Theta$ (the model);

\item[iii)] the observed values of the variables (data) $D\in\mathcal{D}$.
\end{description}

The search space for the posterior probability distribution occurs in the
product space $\mathcal{D\times}\Theta$, and the joint distribution is denoted
as $P\left(  D\text{, }\theta\right)  $. The key idea is going from the old
prior $P_{\text{old}}\left(  \theta\right)  $ to the updated prior
$P_{\text{new}}\left(  \theta\right)  $,%
\begin{equation}
P_{\text{new}}\left(  \theta\right)  \overset{\text{def}}{=}\int
dDP_{\text{new}}\left(  D\text{, }\theta\right)  \text{.}%
\end{equation}
The joint probability $P_{\text{new}}\left(  D\text{, }\theta\right)  $
maximizes the relative entropy functional $S\left[  P\left\vert P_{\text{old}%
}\right.  \right]  $,%
\begin{equation}
S\left[  P\left\vert P_{\text{old}}\right.  \right]  \overset{\text{def}}%
{=}-\int dDd\theta P\left(  D\text{, }\theta\right)  \log\left[
\frac{P\left(  D\text{, }\theta\right)  }{P_{\text{old}}\left(  D\text{,
}\theta\right)  }\right]  \text{,} \label{ef}%
\end{equation}
subject to the given information constraints. Note that $P_{\text{old}}\left(
D\text{, }\theta\right)  $,%
\begin{equation}
P_{\text{old}}\left(  D\text{, }\theta\right)  =P_{\text{old}}\left(
D\left\vert \theta\right.  \right)  P_{\text{old}}\left(  \theta\right)
\text{,}%
\end{equation}
is called here the joint prior, while $P_{\text{old}}\left(  \theta\right)  $
and $P_{\text{old}}\left(  D\left\vert \theta\right.  \right)  $ denote the
standard Bayesian prior and the likelihood, respectively. We emphasize that
both the joint prior and the standard Bayesian prior encode prior information
about $\theta\in\Theta$. Furthermore, despite the fact that the likelihood is
not regarded as prior information in the conventional sense, it will be
considered here as prior information since it represents the a priori
established relation between $\theta\in\Theta$ and $D\in\mathcal{D}$. Let us
specify now the relevant information constraints. First, we impose the
standard normalization constraint,%
\begin{equation}
\int dDd\theta P\left(  D\text{, }\theta\right)  =1\text{.} \label{c1}%
\end{equation}
Second, we consider an information constraint in the form of expected value of
some smooth function $g\left(  \theta\right)  $,%
\begin{equation}
\int dDd\theta g\left(  \theta\right)  P\left(  D\text{, }\theta\right)
\overset{\text{def}}{=}\left\langle g\left(  \theta\right)  \right\rangle
\equiv G\text{.} \label{c2}%
\end{equation}
Finally, we consider the observed data $D^{\prime}$. Within the MrE method,
knowledge of this information leads to an infinite number of constraints,%
\begin{equation}
\int d\theta P\left(  D\text{, }\theta\right)  \equiv P\left(  D\right)
=\delta\left(  D-D^{\prime}\right)  \text{,} \label{c3}%
\end{equation}
for any $D\in\mathcal{D}$ where $\delta$ denotes the Dirac delta function.
Using the Lagrange multipliers technique, we maximize the logarithmic relative
entropy functional in Eq. (\ref{ef}) subject to the constraints in Eqs.
(\ref{c1}), (\ref{c2}), and (\ref{c3}). We impose that the variation with
respect to $P$ of the entropy functional $S\left[  P\left\vert P_{\text{old}%
}\right.  \right]  $ is equal to zero,%
\begin{equation}
\delta\left\{
\begin{array}
[c]{c}%
S\left[  P\left\vert P_{\text{old}}\right.  \right]  +\alpha\left[  \int
dDd\theta P\left(  D\text{, }\theta\right)  -1\right] \\
+\beta\left[  \int dDd\theta g\left(  \theta\right)  P\left(  D\text{, }%
\theta\right)  -G\right] \\
+\int dD\gamma\left(  D\right)  \left[  \int d\theta P\left(  D\text{, }%
\theta\right)  -\delta\left(  D-D^{\prime}\right)  \right]
\end{array}
\right\}  =0\text{.} \label{var}%
\end{equation}
We note that Eq. (\ref{c3}) describes the information in the data and
represents an infinite number of constraints on the family $P\left(  D\text{,
}\theta\right)  $. For this reason, there exist one constraint and one
Lagrange multiplier $\gamma\left(  D\right)  $ for each value of $D$. After
some simple algebra of variations, Eq. (\ref{var}) becomes%
\begin{equation}
\int dDd\theta\left[  -\log P\left(  D\text{, }\theta\right)  -1+\log
P_{\text{old}}\left(  D\text{, }\theta\right)  +\alpha+\beta g\left(
\theta\right)  +\gamma\left(  D\right)  \right]  \delta P\left(  D\text{,
}\theta\right)  =0\text{,} \label{get}%
\end{equation}
for any $\delta P\left(  D\text{, }\theta\right)  $. Therefore, from Eq.
(\ref{get}) we find%
\begin{equation}
P_{\text{new}}\left(  D\text{, }\theta\right)  =P_{\text{old}}\left(  D\text{,
}\theta\right)  e^{-1+\alpha+\beta g\left(  \theta\right)  +\gamma\left(
D\right)  }\text{,} \label{newp}%
\end{equation}
where the Lagrange multipliers $\alpha$, $\beta$, and $\gamma\left(  D\right)
$ can be formally determined by substituting Eq. (\ref{newp}) into Eqs.
(\ref{c1}), (\ref{c2}) and (\ref{c3}), respectively. After some algebra, we
obtain%
\begin{equation}
P_{\text{new}}\left(  D\text{, }\theta\right)  =\frac{e^{\beta g\left(
\theta\right)  }P_{\text{old}}\left(  D\text{, }\theta\right)  \delta\left(
D-D^{\prime}\right)  }{\int d\theta e^{\beta g\left(  \theta\right)
}P_{\text{old}}\left(  D\text{, }\theta\right)  }\text{.} \label{pnew}%
\end{equation}
Observe that the Lagrange multiplier $\beta$ in Eq. (\ref{pnew}) can only be
implicitly determined and depends on the observed data $D^{\prime}$. Finally,
marginalizing $P_{\text{new}}\left(  D\text{, }\theta\right)  $ over the
variable $D$, we obtain the updated prior probability distribution%
\begin{equation}
P_{\text{new}}\left(  \theta\right)  \overset{\text{def}}{=}\int
dDP_{\text{new}}\left(  D\text{, }\theta\right)  =\frac{e^{\beta g\left(
\theta\right)  }P_{\text{old}}\left(  D^{\prime}\text{, }\theta\right)  }{\int
d\theta e^{\beta g\left(  \theta\right)  }P_{\text{old}}\left(  D^{\prime
}\text{, }\theta\right)  }\text{.} \label{gg}%
\end{equation}
For the sake of notational simplicity, define%
\begin{equation}
\xi\left(  D^{\prime}\text{, }\beta\right)  \overset{\text{def}}{=}\int
d\theta e^{\beta g\left(  \theta\right)  }P_{\text{old}}\left(  D^{\prime
}\text{, }\theta\right)  \text{.}%
\end{equation}
Finally, Eq. (\ref{gg}) becomes%
\begin{equation}
P_{\text{new}}\left(  \theta\right)  =P_{\text{old}}\left(  \theta\right)
P_{\text{old}}\left(  D^{\prime}\left\vert \theta\right.  \right)
\frac{e^{\beta g\left(  \theta\right)  }}{\xi\left(  D^{\prime}\text{, }%
\beta\right)  }\text{.} \label{ggg}%
\end{equation}
Note that in the absence of constraints in the form of expected values,
$\beta=0$ and Eq. (\ref{ggg}) reduces to the standard Bayes updating relation%
\begin{equation}
P_{\text{new}}\left(  \theta\right)  =\frac{P_{\text{old}}\left(
\theta\right)  P_{\text{old}}\left(  D^{\prime}\left\vert \theta\right.
\right)  }{P_{\text{old}}\left(  D^{\prime}\right)  }\text{.} \label{bb}%
\end{equation}
For the sake of completeness, we point out that Eq. (\ref{bb}) can be obtained
by combining Bayes theorem,%
\begin{equation}
P_{\text{old}}\left(  \theta\left\vert D\right.  \right)  =\frac
{P_{\text{old}}\left(  \theta\right)  P_{\text{old}}\left(  D\left\vert
\theta\right.  \right)  }{P_{\text{old}}\left(  D\right)  }\text{,}%
\end{equation}
and Bayes rule,%
\begin{equation}
P_{\text{new}}\left(  \theta\right)  =P_{\text{old}}\left(  \theta\left\vert
D^{\prime}\right.  \right)  \text{.}%
\end{equation}
In what follows, we shall regard the MrE method as a tool for approximating
probability distributions.

\subsection{Approximating probability distributions}

For the sake of reasoning, assume that the microstates of a system are labeled
by coordinates $x$. Furthermore, assume that the probability that the system
is in a microstate within a given range $dx$ is described by the canonical
distribution \cite{tseng2008},%
\begin{equation}
P\left(  x\right)  dx=\frac{e^{-\beta\mathcal{H}\left(  x\right)  }%
}{\mathcal{Z}}dx\text{,} \label{p}%
\end{equation}
where the function $Z$ is defined as,%
\begin{equation}
\mathcal{Z}=\int dxe^{-\beta\mathcal{H}\left(  x\right)  }\overset{\text{def}%
}{=}e^{-\beta F}\text{.} \label{ppp}%
\end{equation}
In statistical mechanical terms, the quantity $F$ in Eq. (\ref{ppp}) denotes
the Helmholtz free energy. In general, to describe a system in terms of its
observables, one needs to compute expected values in terms of integrals over
$P$ which, in turn, depends on the Hamiltonian $\mathcal{H}\left(  x\right)
$. Calculating these integrals can be quite challenging especially when the
Hamiltonian exhibits nonlinear interaction terms. To make things more
tractable, one can think of replacing the exact expression of $P$ with an
approximate expression $P_{0}$ provided that the latter preserves the relevant
information content essential for our specific inferences. More specifically,
$P_{0}$ can be identified in two steps. First, identify a suitable family of
trial distributions $\mathcal{F}_{P_{A}}=\left\{  P_{A}\right\}  $ relevant
for our specific problem at hand. Second, select $P_{0}$ from $\mathcal{F}%
_{P_{A}}$. There is no recipe for the first step. The second step is
mechanical. The distribution to be selected is the one that maximizes the
entropy of $P_{A}$ relative to $P$,%
\begin{equation}
S\left[  P_{A}\left\vert P\right.  \right]  =-\int dxP_{A}\left(  x\right)
\log\left[  \frac{P_{A}\left(  x\right)  }{P\left(  x\right)  }\right]
\text{.} \label{re}%
\end{equation}
A very convenient family of trial distributions is given by the canonical
distributions with a modified Hamiltonian $\mathcal{H}_{0}\left(  x\text{;
}\xi\right)  $ where $\xi$ are parameters that label each distribution within
a family,%
\begin{equation}
P_{0}\left(  x\text{; }\xi\right)  dx=\frac{e^{-\beta\mathcal{H}_{0}\left(
x\text{; }\xi\right)  }}{\mathcal{Z}_{0}}dx\text{,} \label{pp}%
\end{equation}
where the $\mathcal{Z}_{0}$ function is given by,%
\begin{equation}
\mathcal{Z}_{0}\overset{\text{def}}{=}e^{-\beta F_{0}\left(  \xi\right)
}=\int dxe^{-\beta\mathcal{H}_{0}\left(  x\text{; }\xi\right)  }\text{.}
\label{ppz}%
\end{equation}
We underline that in Sec. II, the symbol $\theta$ is used to denote one
parameter or many parameters about which one wishes to make inferences. Here,
as mentioned earlier, the parameters $\xi$ are used to label a family of trial
distributions that are canonical with the Hamiltonian\textbf{ }$\mathcal{H}%
_{0}\left(  x\text{; }\xi\right)  $. Substituting Eqs. (\ref{p}), (\ref{ppp}),
(\ref{pp}), and (\ref{ppz}) into Eq. (\ref{re}), we get%
\begin{equation}
S\left[  P_{0}\left\vert P\right.  \right]  =\beta\left(  F-F_{0}\right)
+\beta\left\langle \mathcal{H}_{0}-\mathcal{H}\right\rangle _{0}\text{,}
\label{sp}%
\end{equation}
where,%
\begin{equation}
\left\langle \mathcal{H}_{0}-\mathcal{H}\right\rangle _{0}\overset{\text{def}%
}{=}\int dxP_{0}\left(  x\text{; }\xi\right)  \left(  \mathcal{H}%
_{0}-\mathcal{H}\right)  \text{.}%
\end{equation}
Note that the symbol\textbf{ }$\left\langle \mathcal{\cdot}\right\rangle _{0}%
$\textbf{ }in Eq. (\ref{sp}) denotes the average over the trial
distribution\textbf{ }$P_{0}\left(  x\text{; }\xi\right)  $\textbf{ }and the
subscript\textbf{ }$0$\textbf{ }is simply used to recall that\textbf{ }%
$P_{0}\left(  x\text{; }\xi\right)  $ is canonical with the modified
Hamiltonian\textbf{ }$\mathcal{H}_{0}\left(  x\text{; }\xi\right)  $. Since
$S\left[  P_{0}\left\vert P\right.  \right]  \leq0$, Eq. (\ref{sp}) implies%
\begin{equation}
F\leq F_{0}+\left\langle \mathcal{H}_{0}-\mathcal{H}\right\rangle _{0}\text{.}%
\end{equation}
Loosely speaking, maximizing $S$ is equivalent to minimizing $F_{0}%
+\left\langle \mathcal{H}_{0}-\mathcal{H}\right\rangle _{0}$. This latter
minimization procedure is also known as the Bogoliubov Variational Principle
\cite{kuzemsky} which, as shown here, can be regarded as a special case of the
MrE method.

\begin{table}[ptbh]
\begin{center}%
\begin{tabular}
[c]{|c|c|c|}\hline
FERROMAGNET & SYMBOL & CRITICAL\ TEMPERATURE\\\hline\hline
iron & Fe & 1043\\\hline
cobalt & Co & 1388\\\hline
nickel & Ni & 627\\\hline
gadolinium & Gd & 293\\\hline
dysprosium & Dy & 85\\\hline
\end{tabular}
\end{center}
\caption{A few ferromagnetic materials with corresponding critical
temperatures in degrees Kelvin.}%
\end{table}

\section{Ferromagnetic materials}

One of the most interesting phenomena in solid state physics is represented by
ferromagnetism \cite{huang, mermin, Schroeder}. In spite of the extensive
oversimplifications required to construct a tractable mathematical model
describing such a phenomena, calculations concerning ferromagnetism are among
the most impressive exhibition of brute force computation achieved by
theoretical physicists \cite{onsager, yang}. In what follows, we shall outline
the basics of the so-called Ising model of ferromagnetic materials
\cite{ising}. A few ferromagnetic materials together with their corresponding
critical temperatures are listed in Table I.

\subsection{The Ising Model}

One of the simplest systems that exhibits a transition from ordered to
disordered states is that of a lattice composed of two different types of
objects, $A$ and $B$. Assume that the objects interact with only their nearest
neighbors. If one raises the temperature $T$ of the system at some critical
point $T_{c}$, the system will melt and become completely disordered. A
relatively simple mathematical model of such a system was developed by Ernst
Ising to describe ferromagnetism \cite{ising}. The general expression for the
Ising Hamiltonian is given by \cite{huang},%
\begin{equation}
\mathcal{H}\left\{  s_{i}\right\}  \overset{\text{def}}{=}-\sum_{\left\langle
ij\right\rangle }J_{ij}s_{i}s_{j}-B\sum_{i=1}^{N}s_{i}\text{,} \label{ham}%
\end{equation}
where $s_{i}\in\left\{  +1\text{, }-1\right\}  $ is the spin variable, $N$ is
the cardinality of the spin variables, $\left\{  s_{i}\right\}  $ determines
the spin configuration of the whole system, the symbol $\left\langle
ij\right\rangle $ with $\left\langle ij\right\rangle =\left\langle
ji\right\rangle $ denotes a pair of next neighbor spins, $J_{ij}$ are the
exchange coupling constants, and $B$ is the $z$-component of a uniform
external magnetic field. Ferromagnetic and antiferromagnetic materials are
characterized by exchange coupling coefficients with $J_{ij}>0$ and $J_{ij}%
<0$, respectively, for all pairs $i$, $j$. For instance, iron (Fe) and iron
oxide (FeO) are examples of ferromagnetic and antiferromagnetic materials,
respectively. Their critical temperature is given by $T_{c}^{\left(
Fe\right)  }=1043K$ and $T_{c}^{\left(  FeO\right)  }=198K$ \cite{mermin}, respectively.

For the sake of reasoning, we shall assume in what follows that $B=0$ (zero
external magnetic field) and $J_{ij}=J>0$ for all pairs $i$, $j$ (isotropic
interaction strength). The Hamiltonian in Eq. (\ref{ham}) becomes,%
\begin{equation}
\mathcal{H}\left\{  s_{i}\right\}  \overset{\text{def}}{=}-J\sum_{\left\langle
ij\right\rangle }s_{i}s_{j}\text{.} \label{ham2}%
\end{equation}
The sum over $\left\langle ij\right\rangle $ in Eq. (\ref{ham2}) contains
$\gamma N/2$ terms where $\gamma$ is the number of next nearest neighbors of
each lattice site and depends on the specific type of lattice structure being
considered. For instance, for a two-dimensional square lattice, $\gamma=4$.
The partition function $\mathcal{Z}\left(  T\right)  $ corresponding to the
Hamiltonian in Eq. (\ref{ham2}) is the sum of $2^{N}$ terms and is given by
\cite{huang},%
\begin{equation}
\mathcal{Z}\left(  T\right)  \overset{\text{def}}{=}\sum_{s_{1}}\sum_{s_{2}%
}\text{...}\sum_{s_{N}}e^{-\beta\mathcal{H}\left\{  s_{i}\right\}  }\text{,}
\label{partition}%
\end{equation}
where $\beta=\beta\left(  k_{B}\text{, }T\right)  \overset{\text{def}}%
{=}\left(  k_{B}T\right)  ^{-1}$, with $k_{B}\approx1.38\times10^{-23}JK^{-1}$
denoting the Boltzmann constant. As a side remark, we recall that the
thermodynamic functions (internal energy and thermal capacity, for instance)
can be obtained in the usual manner from the Helmholtz free energy\textbf{
}$F\left(  T\right)  $ \cite{huang},%
\begin{equation}
F\left(  T\right)  \overset{\text{def}}{=}-k_{B}T\log\mathcal{Z}\left(
T\right)  \text{.}%
\end{equation}

\begin{figure}[ptb]
\centering
\includegraphics[width=0.5\textwidth] {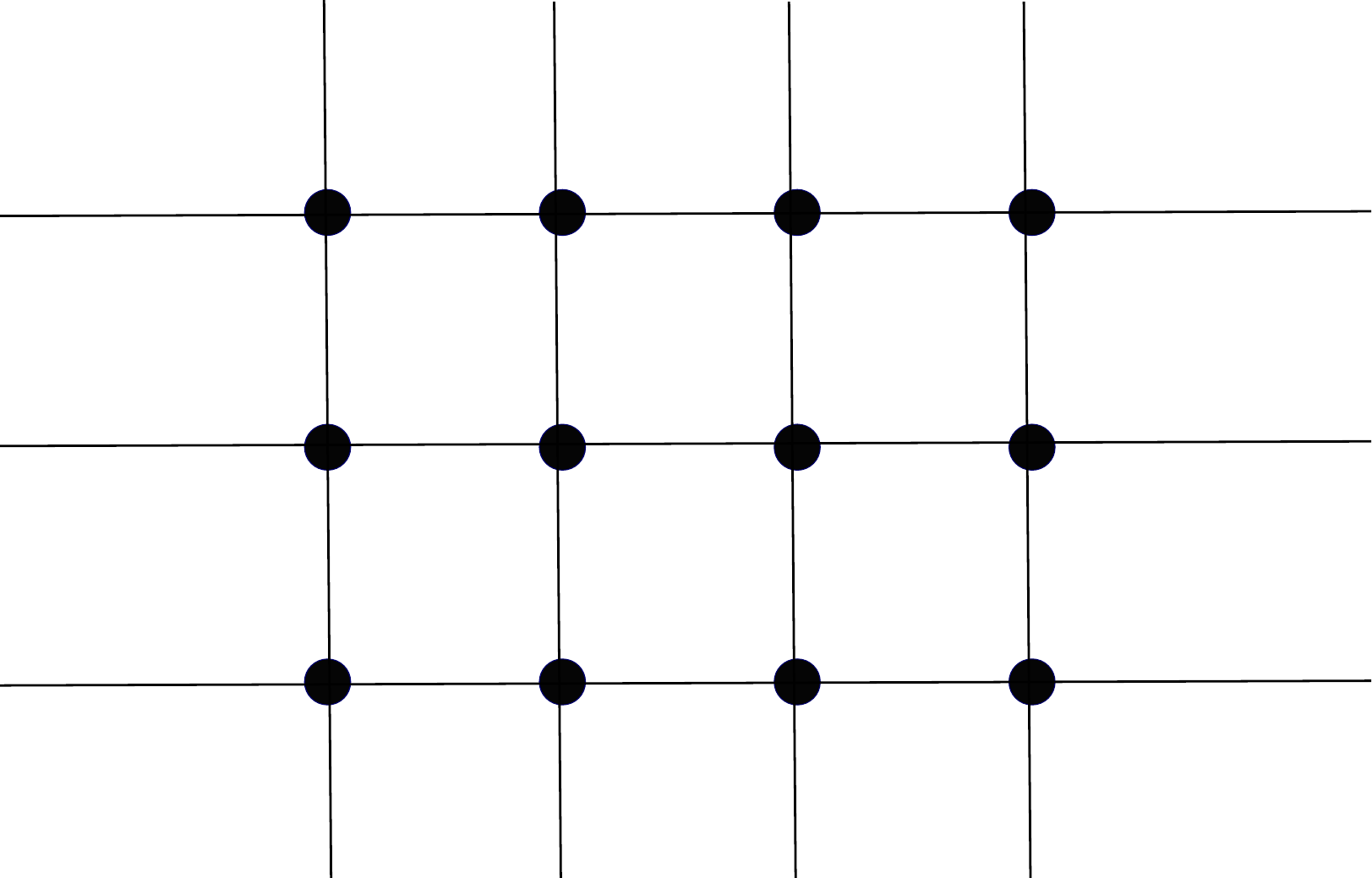}\caption{Schematic illustration of
a rectangular two-dimensional Ising model.}%
\label{fig1}%
\end{figure}

In the two-dimensional case with only nearest neighbor interactions and simple
lattice configurations (square and triangular, for instance), the exact free
energy is known in the hypothesis of zero magnetic field \cite{onsager, yang}.
For the one dimensional case, the partition function $\mathcal{Z}\left(
T\right)  $ can easily be calculated,
\begin{equation}
\mathcal{Z}\left(  T\right)  =2^{N}\left[  \cosh\left(  \beta J\right)
\right]  ^{N-1}\overset{N\gg1}{\approx}\left[  2\cosh\left(  \beta J\right)
\right]  ^{N}\text{.} \label{partition2}%
\end{equation}
Furthermore, the resulting average energy $\mathcal{\bar{H}}$ for $N\gg1$
would then be,
\begin{equation}
\mathcal{\bar{H}}=-\frac{\partial\log Z}{\partial\beta}=-NJ\tanh\left(  \beta
J\right)  ~\text{,} \label{average}%
\end{equation}
where $\tanh\left(  \cdot\right)  $ denotes the hyperbolic tangent function.
For the sake of clarity, a rectangular two-dimensional Ising model is depicted
in FIG. 1.

\subsection{The mean field approximation}

One of the most important starting points for more sophisticated calculations
concerning ferromagnetic transitions is furnished by the so-called mean field
theory (or, molecular field theory \cite{weiss}). The mean field approximation
for the Ising model assumes that each and every spin is subject to a mean
field generated by its next nearest neighbors only \cite{huang}. Such an
approximation turns out to be especially important for three dimensional
materials where each atom may have $6$, $8$ or even $12$ close neighbors
depending on the crystal geometry \cite{Reif, Schroeder}.

We start with the same energy function as given in Eq. (\ref{ham2}) but now
examine only one atom (sometimes called the \emph{central} atom),
\begin{equation}
\mathcal{H}_{i}=-Js_{i}\sum_{j=1}^{n}s_{j}~\text{,}%
\end{equation}
where $\mathcal{H}_{i}$ is the energy ascribed to one atom and $n$ is the
number of nearest neighbors (for instance, a face-centered cubic lattice would
have $n=12$) with atoms at the edge of the material neglected. Assuming that
each neighbor contributes equally and averaging over the neighbor spins around
this one atom, we obtain
\begin{equation}
\mathcal{\bar{H}}_{i}=-Js_{i}n\bar{s}_{j}\text{,}%
\end{equation}
where,%
\begin{equation}
\bar{s}_{j}\overset{\text{def}}{=}\frac{1}{n}\sum_{j=1}^{n}s_{j}\text{,}%
\end{equation}
with $j\in\left\{  1\text{,..., }n\right\}  $. The Boltzmann probability
$p_{i}$ of finding this atom in the $s_{i}$ state is given by,
\begin{equation}
p_{i}=\frac{1}{\mathcal{Z}_{i}}e^{-\beta\mathcal{\bar{H}}_{i}}\text{, }
\label{1}%
\end{equation}
and the partition function $\mathcal{Z}_{i}$ becomes,
\begin{equation}
\mathcal{Z}_{i}=2\cosh\left(  \beta Jn\bar{s}_{j}\right)  \text{.} \label{2}%
\end{equation}
Taking the average of $s_{i}$ for this one atom over its two possible states,
using Eqs. (\ref{1}) and (\ref{2}), the mean magnetization becomes
\begin{equation}
\left\langle s_{i}\right\rangle \overset{\text{def}}{=}\sum_{s_{i}}s_{i}%
~p_{i}=\frac{1}{\mathcal{Z}_{i}}\left[  (1)e^{\beta\mathcal{\bar{H}}_{i}%
}+(-1)e^{-\beta\mathcal{\bar{H}}_{i}}\right]  =\frac{2\sinh\left(  \beta
Jn\bar{s}_{j}\right)  }{2\cosh\left(  \beta Jn\bar{s}_{j}\right)  }%
=\tanh\left(  \beta Jn\bar{s}_{j}\right)  ~\text{.} \label{Expected_M}%
\end{equation}
Furthermore, assuming that \emph{all} atoms will behave like this one labeled
with the $i$ index and assuming that the average spin $\bar{s}$ for the entire
ferromagnetic material is formally defined as,
\begin{equation}
\bar{s}\overset{\text{def}}{=}\left\langle \left\langle s_{i}\right\rangle
\right\rangle =\left\langle \bar{s}_{j}\right\rangle \text{,} \label{as}%
\end{equation}
from Eqs. (\ref{Expected_M}) and (\ref{as}), we find
\begin{equation}
\bar{s}=\tanh\left(  \beta Jn\bar{s}\right)  \text{.} \label{AvgM}%
\end{equation}

\begin{figure}[ptb]
\centering
\includegraphics[width=0.5\textwidth] {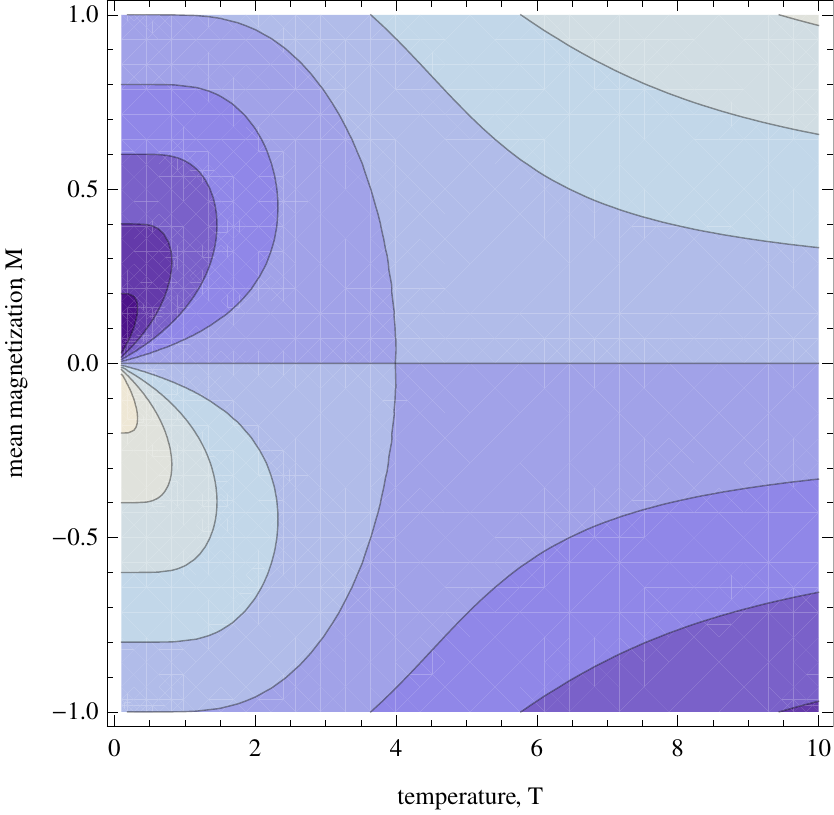}\caption{Contour plot exhibiting
the numerical solution of the mean magnetization $M$ vs. temperature $T$ in
case of the mean field approximation applied to the Ising model. We set
$k_{B}=1$ and assume $n=4$, $J=1$, and the level curves $f\left(  M\text{,
}T\right)  =c$ with $c\in\left\{  -0.8\text{, }-0.6\text{,..., }0\text{,
}0.2\text{,..., }0.6\right\}  . $}%
\label{fig2}%
\end{figure}

Observe that the inner and outer expectation values in the nested expectation
value in Eq. (\ref{as}) used to define the average spin $\bar{s}$ for the
entire ferromagnetic material represent the averages over all possible states
of an atom and over all atoms of the material, respectively. If one considers
Eq. (\ref{ham}) instead of Eq. (\ref{ham2}), Eq. (\ref{AvgM}) can be
generalized and becomes \cite{mac},%
\begin{equation}
\bar{s}=\tanh\left[  \beta\left(  nJ\bar{s}+B\right)  \right]  \text{,}
\label{implicit}%
\end{equation}
where\textbf{ }$B$\textbf{ }denotes the intensity of the applied external
magnetic field. Eq. (\ref{implicit}) can be solved numerically in order to
uncover the mean magnetization $M\overset{\text{def}}{=}\bar{s}$ as a function
of the temperature $T$ of the system. A contour plot of $f\left(  M\text{,
}T\right)  $ defined as,%
\begin{equation}
f\left(  M\text{, }T\right)  \overset{\text{def}}{=}M-\tanh\left[  \frac
{1}{k_{B}T}\left(  nJM+B\right)  \right]  \text{,}%
\end{equation}
as a function of $M$ and $T$ appears in FIG. 2. When an external magnetic
field with strength $B$ is applied, the effective Hamiltonian $\left(
\mathcal{\bar{H}}_{i}\right)  _{\text{eff}}$ for the $i$-th atom becomes,%
\begin{equation}
\left(  \mathcal{\bar{H}}_{i}\right)  _{\text{eff}}=-Js_{i}n\bar{s}_{j}~-\mu
Bs_{i}=-\epsilon s_{i}\text{,} \label{Hamiltonian_effM}%
\end{equation}
where\textbf{ }$\mu\overset{\text{def}}{=}\frac{\epsilon}{B}$\textbf{ }is the
magnetic moment and\textbf{ }$\epsilon\overset{\text{def}}{=}\mu B$\textbf{
}is the net effective energy on the $i$-th atom. As a final remark, we also
point out that if one were to write the full Hamiltonian with the external
field instead of $\mathcal{H}\left\{  s_{i}\right\}  $ as given in Eq.
(\ref{ham2}), then one will not have as simple of a solution for $\mathcal{Z}$
as in Eq. (\ref{partition2}), i.e. $\mathcal{Z}_{\text{eff}}\neq\left[
2\cosh\left(  \beta\epsilon\right)  \right]  ^{N}$.

\section{Maximum relative Entropy method and ferromagnetic materials}

As we have seen in the previous Section, the Ising model and mean field theory
require various assumptions to be fulfilled. Part of the strength of using the
MrE is that one does not need to justify some of the assumptions. Stated
otherwise, one simply supplies the information constraints that one has
available and then allow the method to turn out the least biased solution
based on the information given. This can be demonstrated by using the MrE
method to find an appropriate approximate description of a ferromagnetic material.

Following the procedure outlined in Section II, we begin by determining the
full posterior solution for a ferromagnetic atomic system where we have both
an expectation value constraint as well as observed data. For this example,
the data observed will be the effect of the external magnetic field, $B_{i}$
on each atom. The appropriate entropy to consider becomes,\textbf{ }%
\begin{equation}
S[p|p_{\text{old}}]=-\int d^{N}B\sum_{\left\{  s\right\}  }p\left\{
s,B\right\}  \log\left[  \frac{p\left\{  s,B\right\}  }{p_{\text{old}}\left\{
s,B\right\}  }\right]  \label{H_entropy}%
\end{equation}
where\textbf{ }$p\left\{  s,B\right\}  =p(s_{1},B_{1},$\textbf{\ldots}%
$,s_{N},B_{N})$\textbf{, }$p_{\text{old}}\left\{  s,B\right\}  $\textbf{ }will
be a flat prior (constant in $s$ and uniform in $B$), $N$ denotes the number
of atoms, and%
\begin{equation}
\int d^{N}B\sum_{\left\{  s\right\}  }=\int dB_{1}\ldots dB_{N}\sum_{s_{1}%
}\sum_{s_{2}}\text{...}\sum_{s_{N}}\text{.}%
\end{equation}
Second, we consider the energy constraint in the form of the expectation value
of the Hamiltonian,%
\begin{equation}
\int d^{N}B\sum_{\left\{  s\right\}  }\mathcal{H}\left\{  s\right\}  p\left\{
s,B\right\}  \equiv\left\langle \mathcal{H}\left\{  s\right\}  \right\rangle
\text{ ,} \label{H_constraint}%
\end{equation}
where\textbf{ }$\mathcal{H}\left\{  s\right\}  $\textbf{ }denotes some general
Hamiltonian. For illustrative purposes, we adopt as our starting point the
Ising model so as to show how one can use the MrE to arrive at a similar
approximation as the one produced by mean field theory. Let\textbf{
}$\mathcal{H}\left\{  s\right\}  $\textbf{ }in Eq. (\ref{H_constraint}) be
defined as,\textbf{ }%
\begin{equation}
\mathcal{H}\left\{  s\right\}  \overset{\text{def}}{=}\mathcal{H}%
_{int}+\mathcal{H}_{ext}=-J\sum_{\left\langle ij\right\rangle }s_{i}s_{j}%
-\sum_{i=1}^{N}\mu B_{i}s_{i}\text{.} \label{45}%
\end{equation}
where\textbf{ }$\mathcal{H}_{int}$ equals\textbf{ }$\mathcal{H}$\textbf{ }in
Eq. (\ref{ham2}) and\textbf{ }$\mathcal{H}_{ext}$\textbf{ }is the energy
attributed to the observed external magnetic field\textbf{ }$B_{i}$\textbf{
}acting on the individual atoms. Next is to apply the data constraints. As
mentioned in Section II, there are an infinite number of data constraints
associated with a single data point. Therefore, here we have\textbf{ }%
$N$\textbf{ }sets of them, one set for each data point. For illustrative
purposes, we will assume that the field is uniform and therefore the
value\textbf{ }$B_{i}=B_{j}=B$\textbf{ }for all\textbf{ }$i$, $j\in\left\{
1\text{,..., }N\right\}  $.\textbf{ }We can then simply write the data
constraint as,%
\begin{equation}
\int dB\sum_{\left\{  s\right\}  }p\left\{  s,B\right\}  \equiv p\left\{
B\right\}  =\delta\left(  B-B^{\prime}\right)  \text{.} \label{H_data}%
\end{equation}
Along with the normalization constraint, maximizing the logarithmic relative
entropy with respect to the constraints yields the canonical distribution,%
\begin{equation}
p\left\{  s\right\}  =\frac{1}{\mathcal{Z}^{\prime}}e^{-\beta\mathcal{H}%
^{\prime}\left\{  s\right\}  }\text{,} \label{P}%
\end{equation}
where $p\left\{  s\right\}  =p(s_{1}$\ldots$s_{N})$, $\mathcal{Z}^{\prime}$ is
similar to the partition function from Eq. (\ref{partition}) with the
exception of the data in the Hamiltonian $\mathcal{H}^{\prime}\left\{
s\right\}  $ which is now,%
\begin{equation}
\mathcal{H}^{\prime}\left\{  s\right\}  \overset{\text{def}}{=}\mathcal{H}%
_{int}+\mathcal{H}_{ext}^{\prime}=-J\sum_{\left\langle ij\right\rangle }%
s_{i}s_{j}-\sum_{i=1}^{N}\mu B^{\prime}s_{i}\text{.} \label{Hamiltonian_NF}%
\end{equation}
However, in order to determine the values of quantities such as the critical
temperature, the total magnetic moment and the magnetic susceptibility, we are
faced with the same dilemma as above: we cannot compute the solutions
explicitly. Therefore, we must try to find an approximation that is
computationally tractable. We now wish to use MrE to find an approximation
that is tractable. We accomplish this by first writing down the appropriate
entropy functional $S[p_{\text{A}}|p]$,
\begin{equation}
S[p_{\text{A}}|p]=-\sum_{\left\{  s\right\}  }p_{\text{A}}\log\left[
\frac{p_{\text{A}}}{p}\right]  \text{,} \label{Entropy_ME}%
\end{equation}
where $p$ is the canonical probability distribution as in Eq. (\ref{P}) with
$\mathcal{H}^{\prime}\left\{  s\right\}  $ as given in Eq.
(\ref{Hamiltonian_NF}) and $p_{\text{A}}$ is the approximation that we seek.
We proceed by rewriting the entropy functional $S[p_{\text{A}}|p]$ as,
\begin{equation}
S[p_{\text{A}}|p]=-\sum_{\left\{  s\right\}  }p_{\text{A}}\log p_{\text{A}%
}+\sum_{\left\{  s\right\}  }p_{\text{A}}\log p\text{.} \label{sre}%
\end{equation}
Using Eqs. (\ref{P}) and (\ref{Hamiltonian_NF}), Eq. (\ref{sre}) becomes
\begin{equation}
S[p_{\text{A}}|p]=\frac{1}{k_{B}}S_{\text{A}}-\sum_{\left\{  s\right\}
}p_{\text{A}}\beta\mathcal{H}^{\prime}\left\{  s\right\}  +\beta F\text{,}
\label{sre1}%
\end{equation}
where we have used the fact that the partition function $\mathcal{Z}^{\prime}$
can also be written in terms of the free energy $F$ as $\mathcal{Z}^{\prime
}=e^{-\beta F}$. Eq. (\ref{sre1}) can further be reduced to%
\begin{equation}
S[p_{\text{A}}|p]=\frac{1}{k_{B}}S_{\text{A}}+\beta\left(  F-\left\langle
\mathcal{H}^{\prime}\left\{  s\right\}  \right\rangle _{\text{A}}\right)
\text{,} \label{aa}%
\end{equation}
where $\left\langle \mathcal{H}^{\prime}\left\{  s\right\}  \right\rangle
_{\text{A}}$ can be regarded as the energy $E_{\text{A}}$ of the system and is
formally defined as,%
\begin{equation}
E_{\text{A}}=\left\langle \mathcal{H}^{\prime}\left\{  s\right\}
\right\rangle _{\text{A}}\overset{\text{def}}{=}\sum_{\left\{  s\right\}
}p_{\text{A}}\mathcal{H}^{\prime}\left\{  s\right\}  \text{.} \label{aaa}%
\end{equation}
Since by definition $S[p_{\text{A}}|p]\leq0$, using Eqs. (\ref{aa}) and
(\ref{aaa}) leads to
\begin{equation}
F\leq E_{\text{A}}-TS_{\text{A}}\text{.} \label{Free}%
\end{equation}
When using the MrE method, we maximize the entropy functional in order to find
the best posterior given the information provided. For this specific case, we
have rewritten the entropy in terms of an inequality that compares the free
energy of the system $F$ with the approximate values for the average energy
$E_{\text{A}}$ and the entropy $S_{\text{A}}$. The free energy is minimized to
determine the best approximation for $p_{\text{A}}$. This minimization problem
is also known as the Bogoliubov Variational Principle \cite{Tseng}. Therefore,
from our discussion, we conclude that this Variational Principle is simply a
special case of the MrE method. To be more general, we proceed with using the
MrE method to find the best approximation. As stated earlier, when using the
MrE method, the goal is to search the family of possible posteriors in order
to find the one that maximizes the entropy given the constraints. In addition
to this, we need a solution that is tractable. Therefore, we seek a posterior
$p_{\text{A}}$ that has a form
\begin{equation}
p_{\text{A}}=\frac{1}{\mathcal{Z}_{\text{A}}}e^{-\beta\mathcal{H}_{\text{A}}%
}~\text{,} \label{PA}%
\end{equation}
where $\mathcal{H}_{\text{A}}$ is defined as,
\begin{equation}
\mathcal{H}_{\text{A}}\overset{\text{def}}{=}-\sum_{i}\epsilon_{i\text{A}%
}s_{i}\text{,} \label{ha}%
\end{equation}
and $\epsilon_{i\text{A}}$ is some effective energy similar to $\left(
\mathcal{\bar{H}}_{i}\right)  _{\text{eff}}$ given in Eq.
(\ref{Hamiltonian_effM}). For illustrative purposes, we assume that all atoms
poses the same effective energy so that $\epsilon_{i\text{A}}=\epsilon
_{\text{A}}$ for any $i\in\left\{  1\text{,..., }N\right\}  $. The difference
is that we do not yet know the form of $\epsilon_{\text{A}}$. We continue by
following a similar route to Eq. (\ref{Expected_M}) by writing the expectation
value for $s_{i}$ with respect to $p_{\text{A}}$,
\begin{equation}
\left\langle s_{i}\right\rangle _{\text{A}}=\sum_{\left\{  s_{i}\right\}
}s_{i}~p_{\text{A}}=\tanh\left(  \beta\epsilon_{i\text{A}}\right)
=\tanh\left(  \beta\epsilon_{\text{A}}\right)  ~\text{,} \label{AvgA}%
\end{equation}
except that here we are marginalizing over all atoms except the $i$-th. Notice
that because the effective energy $\epsilon_{\text{A}}$ is constant, the
solution is independent of the index $i$. Next we maximize $S[p_{\text{A}}|p]$
in\ Eq. (\ref{Entropy_ME}) or, in keeping with the current case, we minimize
the free energy $F$. From Eq. (\ref{aa}), we have
\begin{equation}
\beta F_{\min}=-\frac{1}{k_{B}}S_{\text{A}}+\beta\left\langle \mathcal{H}%
^{\prime}\left\{  s\right\}  \right\rangle _{\text{A}}\text{.} \label{a4}%
\end{equation}
Substituting Eq. (\ref{PA}) into Eq. (\ref{a4}), after some algebra we get%
\begin{equation}
F_{\min}=\left(  \frac{1}{\beta}\sum_{\left\{  s\right\}  }p_{\text{A}}\log
e^{-\beta\mathcal{H}_{\text{A}}}-\frac{1}{\beta}\sum_{\left\{  s\right\}
}p_{\text{A}}\log\mathcal{Z}_{\text{A}}\right)  +\sum_{\left\{  s\right\}
}p_{\text{A}}\mathcal{H}^{\prime}\left\{  s\right\}  ~\text{.} \label{pull}%
\end{equation}
Note that $\mathcal{Z}_{\text{A}}$ is a constant and can be extracted from the
summation in Eq. (\ref{pull}). Furthermore, since $\sum_{\left\{  s\right\}
}p_{\text{A}}=1$, Eq. (\ref{pull}) becomes
\begin{equation}
F_{\min}=\left(  -\frac{1}{\beta}\sum_{\left\{  s\right\}  }p_{\text{A}}%
\beta\mathcal{H}_{\text{A}}-\frac{1}{\beta}\log\mathcal{Z}_{\text{A}}\right)
+\sum_{\left\{  s\right\}  }p_{\text{A}}\mathcal{H}^{\prime}\left\{
s\right\}  \text{,}%
\end{equation}
or expressed otherwise,%
\begin{equation}
F_{\min}=-\left\langle \mathcal{H}_{\text{A}}\right\rangle _{\text{A}}%
-\frac{1}{\beta}\log\mathcal{Z}_{\text{A}}+\left\langle \mathcal{H}^{\prime
}\left\{  s\right\}  \right\rangle _{\text{A}}~\text{.} \label{fmin}%
\end{equation}
We now substitute the explicit expressions for the Hamiltonians $\mathcal{H}%
^{\prime}\left\{  s\right\}  $ in Eq. (\ref{Hamiltonian_NF}) and
$\mathcal{H}_{\text{A}}$ in Eq. (\ref{ha}) into the function $F_{\min}$ in Eq.
(\ref{fmin}). After some algebra, \ we find
\begin{equation}
F_{\min}=-\left\langle -\epsilon_{\text{A}}\sum_{i}s_{i}\right\rangle
_{\text{A}}-\frac{1}{\beta}\log\mathcal{Z}_{\text{A}}+\left\langle
-J\sum_{i,j}^{N,n}s_{i}s_{j}-\sum_{i}^{N}\mu B^{\prime}s_{i}\right\rangle
_{\text{A}}~\text{.} \label{fmin1}%
\end{equation}
Noting that,%
\begin{equation}
\left\langle \sum_{i}s_{i}\right\rangle _{\text{A}}=\sum_{i}\left\langle
s_{i}\right\rangle _{\text{A}}=N\left\langle s_{i}\right\rangle _{\text{A}%
}=N\tanh\left(  \beta\epsilon_{\text{A}}\right)  \text{,}%
\end{equation}
$F_{\min}$ \ becomes,
\begin{equation}
F_{\min}=\epsilon_{\text{A}}N\tanh\left(  \beta\epsilon_{\text{A}}\right)
-\frac{1}{\beta}\log\mathcal{Z}_{\text{A}}-J\sum_{i,j}^{N,n}\left\langle
s_{i}s_{j}\right\rangle _{\text{A}}-\mu B^{\prime}N\tanh\left(  \beta
\epsilon_{\text{A}}\right)  ~\text{.} \label{fmin2}%
\end{equation}
Substituting $\mathcal{Z}_{\text{A}}\overset{N\gg1}{\approx}\left[
2\cosh\left(  \beta\epsilon_{\text{A}}\right)  \right]  ^{N}$ into Eq.
(\ref{fmin2}) yields,
\begin{equation}
F_{\min}=\epsilon_{\text{A}}N\tanh\left(  \beta\epsilon_{\text{A}}\right)
-\frac{1}{\beta}\log\left[  2\cosh\left(  \beta\epsilon_{\text{A}}\right)
\right]  ^{N}-J\sum_{i,j}^{N,n}\left\langle s_{i}s_{j}\right\rangle
_{\text{A}}-\mu B^{\prime}N\tanh\left(  \beta\epsilon_{\text{A}}\right)
\text{.}%
\end{equation}
From Eq. (\ref{AvgA}), $\left\langle s_{i}\right\rangle _{\text{A}}%
=\tanh\left(  \beta\epsilon_{\text{A}}\right)  $ and does not depend on the
index $i$, so that%
\begin{equation}
\left\langle s_{i}\right\rangle _{\text{A}}=\left\langle s_{j}\right\rangle
_{\text{A}}\text{, and }\left\langle s_{i}s_{j}\right\rangle _{\text{A}%
}=\left\langle s_{i}\right\rangle _{\text{A}}\left\langle s_{j}\right\rangle
_{\text{A}}\text{.}%
\end{equation}
Therefore, $F_{\min}$ can be rewritten as
\begin{equation}
F_{\min}=\epsilon_{\text{A}}N\tanh\left(  \beta\epsilon_{\text{A}}\right)
-\frac{1}{\beta}N\log\left[  2\cosh\left(  \beta\epsilon_{\text{A}}\right)
\right]  -J\sum_{i,j}^{N,n}\left\langle s_{i}\right\rangle _{\text{A}%
}\left\langle s_{j}\right\rangle _{\text{A}}-\mu B^{\prime}N\tanh\left(
\beta\epsilon_{\text{A}}\right)  ~\text{.} \label{mmm}%
\end{equation}
Substituting Eq. (\ref{AvgA}) into Eq. (\ref{mmm}) yields,
\begin{equation}
F_{\min}=\epsilon_{\text{A}}N\tanh\left(  \beta\epsilon_{\text{A}}\right)
-\frac{N}{\beta}\log\left[  2\cosh\left(  \beta\epsilon_{\text{A}}\right)
\right]  -J\frac{1}{2}Nn\left[  \tanh\left(  \beta\epsilon_{\text{A}}\right)
\right]  ^{2}-\mu B^{\prime}N\tanh\left(  \beta\epsilon_{\text{A~}}\right)
\text{,} \label{minimi}%
\end{equation}
where the factor $n$ specifies the number of nearest neighbors, the factor $N$
is from the total number of atoms and the $1/2$ appears in order to take into
account double counting. We now minimize $F_{\min}$ in Eq. (\ref{minimi}) with
respect to $\epsilon_{\text{A}}$,
\begin{align}
\frac{\partial F_{\min}}{\partial\epsilon_{\text{A}}}  &  =0=N\tanh\left(
\beta\epsilon_{\text{A}}\right)  +N\epsilon_{\text{A}}\frac{\beta}{\cosh
^{2}\left(  \beta\epsilon_{\text{A}}\right)  }-\frac{N}{\beta}\frac{\beta
\sinh\left(  \beta\epsilon_{\text{A}}\right)  }{\cosh\left(  \beta
\epsilon_{\text{A}}\right)  }+\nonumber\\
&  -J\frac{1}{2}Nn2\tanh\left(  \beta\epsilon_{\text{A}}\right)  \frac{\beta
}{\cosh^{2}\left(  \beta\epsilon_{\text{A}}\right)  }-\mu B^{\prime}%
N\frac{\beta}{\cosh^{2}\left(  \beta\epsilon_{\text{A}}\right)  }~\text{,}%
\end{align}
that is,
\begin{align}
0  &  =N\tanh\left(  \beta\epsilon_{\text{A}}\right)  +N\epsilon_{\text{A}%
}\frac{\beta}{\cosh^{2}\left(  \beta\epsilon_{\text{A}}\right)  }%
-N\tanh\left(  \beta\epsilon_{\text{A}}\right)  +\nonumber\\
&  -JNn\tanh\left(  \beta\epsilon_{\text{A}}\right)  \frac{\beta}{\cosh
^{2}\left(  \beta\epsilon_{\text{A}}\right)  }-\mu B^{\prime}N\frac{\beta
}{\cosh^{2}\left(  \beta\epsilon_{\text{A}}\right)  }~\text{.} \label{simply}%
\end{align}
After canceling terms in Eq. (\ref{simply}), we have
\begin{equation}
0=\epsilon_{\text{A}}-\mu B^{\prime}-Jn\tanh\beta\epsilon_{\text{A}}%
\end{equation}
that is,
\begin{equation}
\epsilon_{\text{A}}-\mu B^{\prime}=Jn\tanh\left(  \beta\epsilon_{\text{A}%
}\right)  ~\text{.} \label{EA}%
\end{equation}
Eq. (\ref{EA}) is our final result for this Section. In FIG. 2, the numerical
solution of the mean magnetization vs. temperature is reported. We note that
for $T=T_{c}=4$ and $h=0$\textbf{,}%
\begin{equation}
\lim_{T\rightarrow T_{c}}\left\vert \frac{dM\left(  T\right)  }{dT}\right\vert
=\infty\text{.}%
\end{equation}
The critical temperature $T_{c}$ is the temperature at which spontaneous
magnetization in a lattice of magnetic material begins to appear, and this is
where a phase transition occurs \cite{stanleybook}. Roughly speaking, a phase
transition in a lattice is a singularity in the limit of $\frac{\log Z\left(
T\right)  }{D}$ as $D$, the size of the lattice, approaches infinity. For the
sake of completeness, we remark here that while Onsager was the first to
obtain a closed-form solution to the Ising two-dimensional ferromagnetic model
in the absence of an external magnetic field \cite{onsager}, Yang was the
first to publish the exact calculation of the spontaneous magnetization for a
two-dimensional Ising model \cite{yang}. Returning to FIG. 2, we note that for
$T>T_{c}$, there is only one approximating probability distribution that is
uniform over all states. For $T<T_{c}$, there are two minima that correspond
to approximating distributions that are symmetry-broken, with all spins more
likely to be down, or all spins more likely to be up. Furthermore, notice that
if $B^{\prime}=0$ in Eq. (\ref{EA}), we recover our solution using the mean
field approximation above. As a matter of fact, letting $\epsilon_{\text{A}%
}=Jn\bar{s}$, we rewrite Eq. (\ref{EA}) as,
\begin{equation}
Jn\bar{s}=Jn\tanh\left(  \beta Jn\bar{s}\right)  ~\text{,}%
\end{equation}
that is, we reattain Eq. (\ref{AvgM}),%
\begin{equation}
\bar{s}=\tanh\left(  \beta Jn\bar{s}\right)  ~\text{.}%
\end{equation}
However, from Eq. (\ref{EA})\ alone we can formally solve for the critical
temperature, the total magnetic moment and the magnetic susceptibility (that
is, the ratio between the magnetization of the material and the strength of
the magnetic field applied to the material) in the usual way \cite{Caticha}.
This is true even though we still do not know the explicit form for
$\epsilon_{\text{A}}$. Using this MrE approach, we did not need to assume that
all atoms behave like a central atom and we did not need to know the explicit
form of the effective energy $\epsilon_{\text{A}}$, only that there was one.
Following the MrE method we simply processed all of the information that we
had available. Notice that we also no longer need to assume Ising conditions.
We further explore such considerations in the next Section.

\section{Example toy applications}

In this Section, we employ MrE techniques to infer the numerical estimates of
effective energy levels of atoms in two cases: i) atoms with three possible
states; ii) defective atoms in a crystal lattice.

\subsection{Atoms with three possible states}

To illustrate the use of MrE in determining critical temperatures, we look at
the Ising model with three possible states: spin up, spin down, and no spin.
In this example, we will let $s\in\{+1,0,-1\}$ where an atom in the $0$-state
would contribute no energy. We follow the same line of reasoning outlined in
the previous Section up until Eq. (\ref{AvgA}) where the new expected value
for the $i$-th spin variable $\left\langle s_{i}\right\rangle _{\text{A}}$ is
now given by,%
\begin{equation}
\left\langle s_{i}\right\rangle _{\text{A}}\overset{\text{def}}{=}%
\sum_{\left\{  s_{i}\right\}  }s_{i}~p_{\text{A}}=\frac{2\sinh\left(
\beta\epsilon_{\text{A}}\right)  }{2\cosh\left(  \beta\epsilon_{\text{A}%
}\right)  +1}\text{~.}%
\end{equation}
After some tedious algebra, the new expression for $F_{\min}$ becomes
\begin{equation}
F_{\min}=\epsilon_{\text{A}}N\frac{2\sinh\left(  \beta\epsilon_{\text{A}%
}\right)  }{2\cosh\left(  \beta\epsilon_{\text{A}}\right)  +1}-\frac{N}{\beta
}\log\left[  2\cosh\left(  \beta\epsilon_{\text{A}}\right)  +1\right]
-J\frac{1}{2}Nn\left[  \frac{2\sinh\left(  \beta\epsilon_{\text{A}}\right)
}{2\cosh\left(  \beta\epsilon_{\text{A}}\right)  +1}\right]  ^{2}-\mu
B^{\prime}N\frac{2\sinh\left(  \beta\epsilon_{\text{A}}\right)  }%
{2\cosh\left(  \beta\epsilon_{\text{A}}\right)  +1}~\text{.}%
\end{equation}

Minimizing this function $F_{\min}$ once again with respect to $\epsilon
_{\text{A}}$ yields,%
\begin{align}
\left.  \frac{\partial F_{\min}}{\partial\epsilon_{\text{A}}}\right\vert
_{\epsilon_{\text{A}}=~\epsilon_{\text{A}\min}}  &  =0=N\frac{2\sinh\left(
\beta\epsilon_{\text{A}}\right)  }{2\cosh\left(  \beta\epsilon_{\text{A}%
}\right)  +1}+N\epsilon_{\text{A}}\left(  \frac{2\beta\cosh\left(
\beta\epsilon_{\text{A}}\right)  }{2\cosh\left(  \beta\epsilon_{\text{A}%
}\right)  +1}-\frac{4\beta\sinh^{2}\left(  \beta\epsilon_{\text{A}}\right)
}{\left[  2\cosh\left(  \beta\epsilon_{\text{A}}\right)  +1\right]  ^{2}%
}\right)  +\nonumber\\
& \nonumber\\
&  -\frac{N}{\beta}\frac{2\beta\sinh\left(  \beta\epsilon_{\text{A}}\right)
}{2\cosh\left(  \beta\epsilon_{\text{A}}\right)  +1}-J\frac{1}{2}Nn2\left(
\frac{2\sinh\left(  \beta\epsilon_{\text{A}}\right)  }{2\cosh\left(
\beta\epsilon_{\text{A}}\right)  +1}\right)  \left(  \frac{2\beta\cosh\left(
\beta\epsilon_{\text{A}}\right)  }{2\cosh\left(  \beta\epsilon_{\text{A}%
}\right)  +1}-\frac{4\beta\sinh^{2}\left(  \beta\epsilon_{\text{A}}\right)
}{\left[  2\cosh\left(  \beta\epsilon_{\text{A}}\right)  +1\right]  ^{2}%
}\right)  +\nonumber\\
& \nonumber\\
&  -\mu B^{\prime}N\left(  \frac{2\beta\cosh\left(  \beta\epsilon_{\text{A}%
}\right)  }{2\cosh\left(  \beta\epsilon_{\text{A}}\right)  +1}-\frac
{4\beta\sinh^{2}\left(  \beta\epsilon_{\text{A}}\right)  }{\left[
2\cosh\left(  \beta\epsilon_{\text{A}}\right)  +1\right]  ^{2}}\right)
~\text{.} \label{mess}%
\end{align}
After canceling terms in Eq. (\ref{mess}), we arrive at
\begin{equation}
\epsilon_{\text{A}}-\mu B^{\prime}=Jn\left(  \frac{2\sinh\left(  \beta
\epsilon_{\text{A}}\right)  }{2\cosh\left(  \beta\epsilon_{\text{A}}\right)
+1}\right)  \text{.} \label{cc}%
\end{equation}

\begin{figure}[ptb]
\centering
\includegraphics[width=0.5\textwidth] {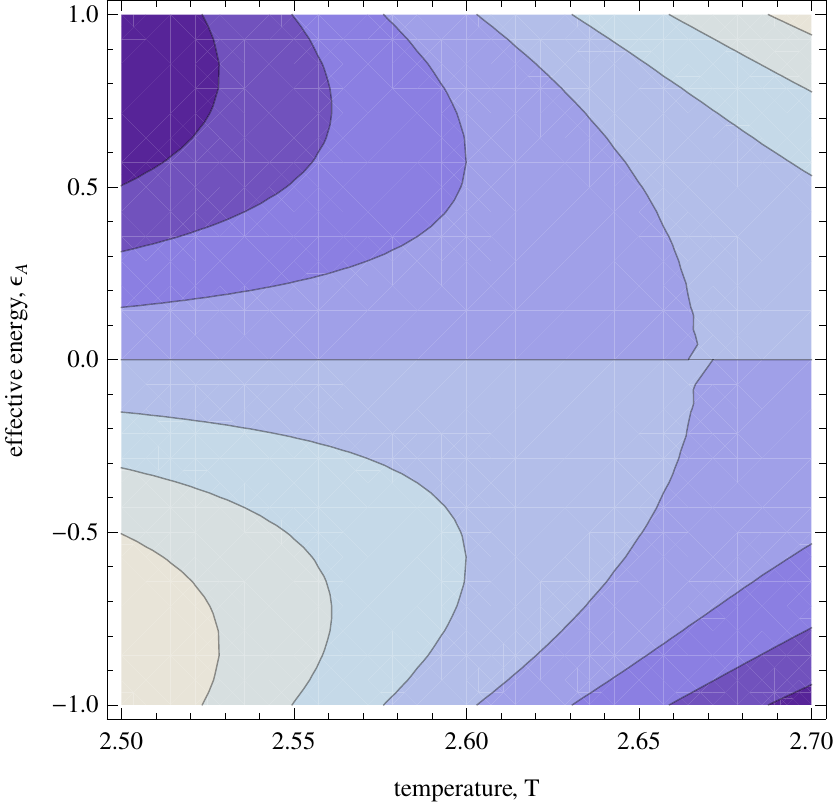}\caption{Contour plot exhibiting
the numerical solution of the effective energy $\epsilon_{A}$ vs. temperature
$T$. We set $k_{B}=1$ and assume $n=4$, $J=1$, $\mu=1$, $B^{\prime}=1$ and the
level curves $f\left(  \epsilon_{A}\text{, }T\right)  =c$ with $c\in\left\{
-1.03\text{, }-1.02\text{, }-1.01\text{, }-1\text{, }-0.99\text{,
}-0.98\text{, }-0.97\right\}  $.}%
\label{fig3}%
\end{figure}

Eq. (\ref{cc}) is our final result. A contour plot of $f\left(  \epsilon
_{\text{A}}\text{, }T\right)  $ defined as,%
\begin{equation}
f\left(  \epsilon_{\text{A}}\text{, }T\right)  \overset{\text{def}}{=}%
\epsilon_{\text{A}}-\mu B^{\prime}-Jn\left(  \frac{2\sinh\left(
\frac{\epsilon_{\text{A}}}{k_{B}T}\right)  }{2\cosh\left(  \frac
{\epsilon_{\text{A}}}{k_{B}T}\right)  +1}\right)  \text{,}%
\end{equation}
as a function of $\epsilon_{\text{A}}$ and $T$ appears in FIG. 3. As stated
earlier, this equation can then be used as above to solve for the critical
temperature, the total magnetic moment and the magnetic susceptibility for
this specific ferromagnetic material. As a side remark, we note that to make
this result a little more general we can also write,
\begin{equation}
\epsilon_{\text{A}}-\mu B^{\prime}=Jn_{\text{G}}\left(  -\frac{\partial
\ln\mathcal{Z}_{\text{A}}}{\partial\epsilon_{\text{A}}}\right)  =Jn_{\text{G}%
}\beta\left(  \frac{\partial F_{\text{A}}}{\partial\epsilon_{\text{A}}%
}\right)  =Jn_{\text{G}}\left(  \left\langle s_{i}\right\rangle _{\text{A}%
}\right)  ~\text{,} \label{EA_G}%
\end{equation}
where $\mathcal{Z}_{\text{A}}$ is once again the approximate partition
function, $n_{\text{G}}$ is the number of nearest neighbors and $F_{\text{A}}$
is the approximate free energy.\textbf{ }Despite its lack of elegance, our
entropic analysis leading to Eq. (\ref{cc}) together with FIG. 3 seems to
support the idea that the introduction of additional (allowed) states in the
atomic ferromagnetic structure is consistent with both space shifts and
nonhomogeneities of the critical temperature \cite{jean00, kallin08}. A deeper
understanding of these specific aspects require a deeper analysis that we
leave to future investigations.

As a side remark, although unnecessary and perhaps impractical, it should be
noted that the\textbf{ }\textit{spin state }$0$ could be introduced as an
observable quantity in a way that is analogous to how $B$ in Eq.
(\ref{H_data}) was implemented.

\subsection{Defects in a crystal lattice}

In \cite{ashkin1943}, Onsager's method was used to study the physical
properties of a two-dimensional square lattice containing four kinds of atoms
under the assumption that only the interaction of nearest neighbors was
important and that only two distinct energies of interaction were permitted.
Here, following this line of investigation, we examine the case where one of
the \textit{atoms} is actually a missing one. In real cases, it would be very
difficult to know how many atoms there are of each type. Indeed, the point of
the \textit{central atom} idea is that we cannot know the states of all the
atoms in a lattice, so we must assume they are all the same. However, it can
be that when a crystal is scanned, the scan can indicate where there might be
impurities or defects. In solid state physics, a point defect (or, vacancy) in
a monatomic Bravais lattice occurs whenever a lattice site that would usually
be occupied by an ion in the perfect crystal has not any ion associated with
it. For the sake of clarity, a vacancy is illustrated in FIG. 4. In the
previous application, we examined a three state atom. In this next example, we
shall examine a material that has a known defect or defects. In this case, we
need to use\ two effective energy terms. One for the atoms that are surrounded
by non-defective atoms, $\epsilon_{\text{ND}}$ and another for the ones
affected by the defect, $\epsilon_{\text{D}}$. For illustrative purposes, we
will examine the case where we only have one defective atom. This means that
there are $N-n-1$ atoms that are surrounded by non-defective atoms, $n$ atoms
that have one defective atom next to it and $1$ atom which is the defective
atom. For our purposes, let us think of the defect as an empty slot. Given
these conditions, the actual Hamiltonian $\mathcal{H}\left\{  s\right\}  $ of
the system is given by,%
\begin{equation}
\mathcal{H}\left\{  s\right\}  =\mathcal{H}_{int}+\mathcal{H}_{ext}%
=-J\sum_{i,j}^{N-n-1,n}s_{i}s_{j}-J\sum_{k,l}^{n,n-1}s_{k}s_{l}-Js_{0}%
-\sum_{i}^{N-n-1}\mu B_{i}s_{i}-\sum_{k}^{n}\mu B_{i}s_{k}-\mu B_{i}%
s_{0}~\text{,} \label{himpo}%
\end{equation}
where $i$ labels the non-defective atoms, $j$ labels the neighbors ($n$) for
these atoms, $k$ are the atoms affected by the defect, $l$ are the neighbors
($n-1)$ of the affected atoms and $s_{0}$ is the spin variable of the
defective atom. Since we are looking at this defect as an empty slot, we let
$s_{0}=0$. Now we write our estimated Hamiltonian $\mathcal{H}_{\text{A}}$ for
this case as,
\begin{equation}
\mathcal{H}_{\text{A}}=-\epsilon_{\text{ND}}\sum_{i}s_{i}-\epsilon_{\text{D}%
}\sum_{k}s_{k}~\text{.} \label{himpa}%
\end{equation}

\begin{figure}[ptb]
\centering
\includegraphics[width=0.3\textwidth] {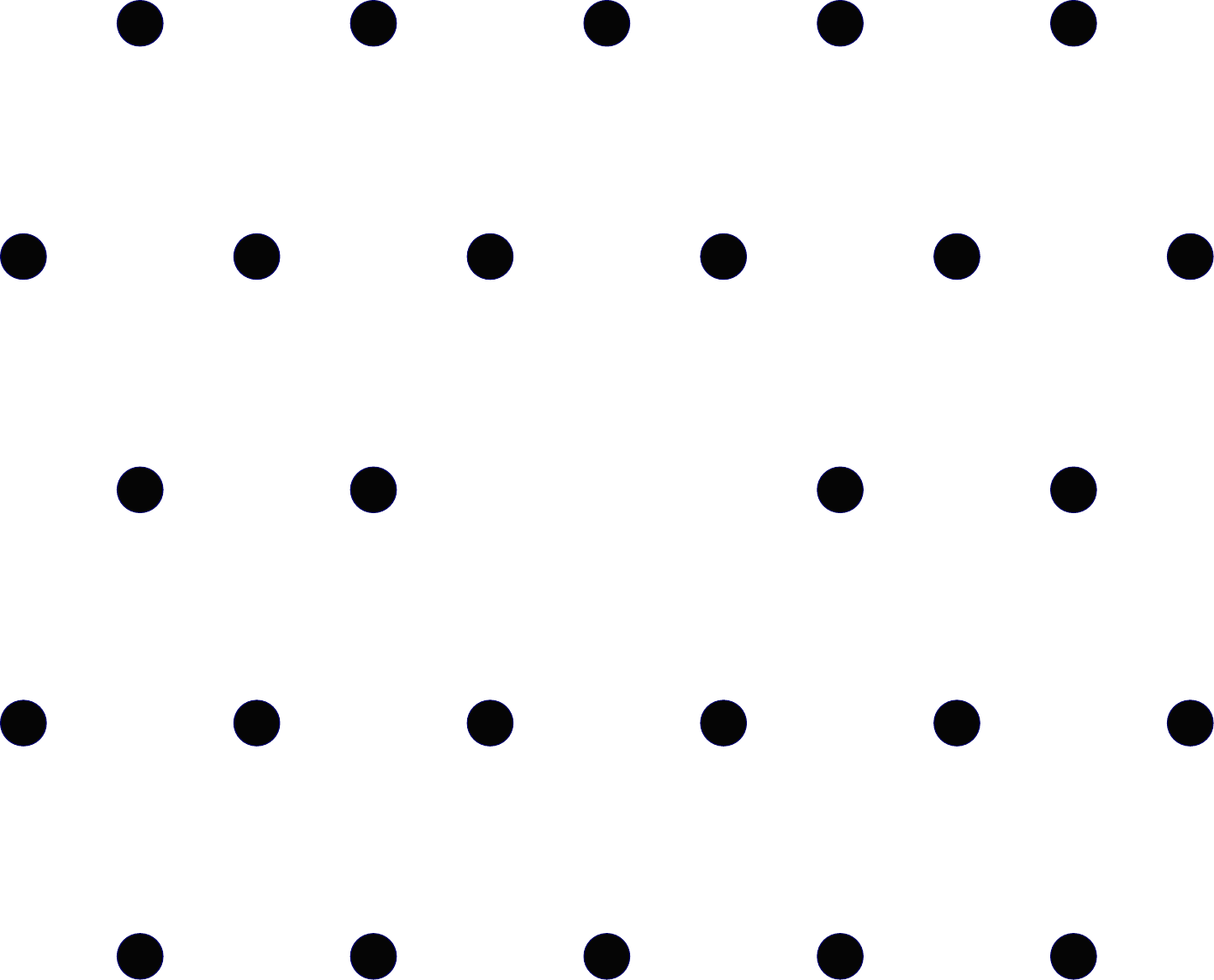}\caption{Schematic illustration of
a point defect (vacancy) in a monatomic Bravais lattice.}%
\label{fig4}%
\end{figure}Following the same procedures outlined above, we attain two
expected values for the spin variables, one for each effective energy,
\begin{equation}
\left\langle s_{i}\right\rangle _{\text{A}}\overset{\text{def}}{=}%
\sum_{\left\{  s_{i}\right\}  }s_{i}~p_{\text{A}}=\tanh\left(  \beta
\epsilon_{\text{ND}}\right)  \text{,}~ \label{S_ND}%
\end{equation}
and,%
\begin{equation}
\left\langle s_{k}\right\rangle _{\text{A}}\overset{\text{def}}{=}%
\sum_{\left\{  s_{i}\right\}  }s_{k}~p_{\text{A}}=\tanh\left(  \beta
\epsilon_{\text{D}}\right)  ~\text{.} \label{S_D}%
\end{equation}
Once again, we write down the function $F_{\min}$ we wish to minimize as%
\begin{equation}
F_{\min}=-\left\langle \mathcal{H}_{\text{A}}\right\rangle _{\text{A}}%
-\frac{1}{\beta}\log\mathcal{Z}_{\text{A}}+\left\langle \mathcal{H}^{\prime
}\left\{  s\right\}  \right\rangle _{\text{A}}~\text{.} \label{fimpo}%
\end{equation}
where $\mathcal{H}^{\prime}\left\{  s\right\}  $ is arrived at in a similar
way to Eq. (\ref{Hamiltonian_NF}). Substituting Eqs. (\ref{himpo}) and
(\ref{himpa}) into Eq. (\ref{fimpo}) for $F_{\min}$ yields,
\begin{align}
F_{\min}  &  =-\left\langle -\epsilon_{\text{ND}}\sum_{i}^{N-n-1}%
s_{i}-\epsilon_{\text{D}}\sum_{k}^{n}s_{k}\right\rangle _{\text{A}}-\frac
{1}{\beta}\log\mathcal{Z}_{\text{A}}+\nonumber\\
& \nonumber\\
&  +\left\langle -J\sum_{i,j}^{N-n-1,n}s_{i}s_{j}-J\sum_{k,l}^{n,n-1}%
s_{k}s_{l}-\sum_{i}^{N-n-1}\mu B^{\prime}s_{i}-\sum_{k}^{n}\mu B^{\prime}%
s_{k}\right\rangle _{\text{A}}~\text{.} \label{car}%
\end{align}
Observe that,%
\begin{equation}
\left\langle \sum_{i}s_{i}\right\rangle _{\text{A}}=\sum_{i}\left\langle
s_{i}\right\rangle _{\text{A}}=\left(  N-n-1\right)  \left\langle
s_{i}\right\rangle _{\text{A}}=\left(  N-n-1\right)  \tanh\left(
\beta\epsilon_{\text{ND}}\right)  \text{,} \label{asss}%
\end{equation}
and,
\begin{equation}
\left\langle \sum_{k}s_{k}\right\rangle _{\text{A}}=\sum_{k}\left\langle
s_{k}\right\rangle _{\text{A}}=n\left\langle s_{k}\right\rangle _{\text{A}%
}=n\tanh\left(  \beta\epsilon_{\text{D}}\right)  \text{.} \label{assss}%
\end{equation}
Substituting Eqs. (\ref{asss}) and (\ref{assss}) into Eq. (\ref{car}), after
some algebra, we obtain
\begin{align}
F_{\min}  &  =\epsilon_{\text{ND}}\left(  N-n-1\right)  \tanh\left(
\beta\epsilon_{\text{ND}}\right)  +\epsilon_{\text{D}}n\tanh\left(
\beta\epsilon_{\text{D}}\right)  -\frac{1}{\beta}\log\left(  \mathcal{Z}%
_{\text{ND}}\mathcal{Z}_{\text{D}}\right)  +\nonumber\\
& \nonumber\\
&  -J\sum_{i,j}^{N-n-1,n}\left\langle s_{i}s_{j}\right\rangle _{\text{ND}%
}-J\sum_{k,l}^{n,n-1}\left\langle s_{k}s_{l}\right\rangle _{\text{D}}-\mu
B^{\prime}\left[  \epsilon_{\text{ND}}\left(  N-n-1\right)  \tanh\left(
\beta\epsilon_{\text{ND}}\right)  +\epsilon_{\text{D}}n\tanh\left(
\beta\epsilon_{\text{D}}\right)  \right]  ~\text{,}%
\end{align}
where $\mathcal{Z}_{\text{A}}\overset{\text{def}}{=}\mathcal{Z}_{\text{ND}%
}\mathcal{Z}_{\text{D}}$. Substituting $Z_{\text{ND}}\overset{N\gg1}{\approx
}\left[  2\cosh\left(  \beta\epsilon_{\text{ND}}\right)  \right]  ^{\left(
N-n-1\right)  }$ and $Z_{\text{D}}\overset{N\gg1}{\approx}\left[
2\cosh\left(  \beta\epsilon_{\text{D}}\right)  \right]  ^{\left(  n\right)  }$
yields,
\begin{align}
F_{\min}  &  =\epsilon_{\text{ND}}\left(  N-n-1\right)  \tanh\left(
\beta\epsilon_{\text{ND}}\right)  +\epsilon_{\text{D}}n\tanh\left(
\beta\epsilon_{\text{D}}\right)  -\frac{1}{\beta}\log\left[  2\cosh\left(
\beta\epsilon_{\text{ND}}\right)  \right]  ^{N-n-1}+\nonumber\\
& \nonumber\\
-\frac{1}{\beta}\log\left[  2\cosh\left(  \beta\epsilon_{\text{D}}\right)
\right]  ^{n-1}  &  -J\sum_{i,j}^{N-n-1,n}\left\langle s_{i}s_{j}\right\rangle
_{\text{ND}}-J\sum_{k,l}^{n,n-1}\left\langle s_{k}s_{l}\right\rangle
_{_{\text{D}}}-\mu B^{\prime}\left[  \epsilon_{\text{ND}}\left(  N-n-1\right)
\tanh\left(  \beta\epsilon_{\text{ND}}\right)  +\epsilon_{\text{D}}%
n\tanh\left(  \beta\epsilon_{\text{D}}\right)  \right]  ~\text{.}%
\end{align}
Since $\left\langle s_{i}\right\rangle _{\text{A}}$ in Eq. (\ref{S_ND}) and
$\left\langle s_{k}\right\rangle _{\text{A}}$ in Eq. (\ref{S_D}) are
independent of $i$ and $k$, respectively, we have $\left\langle s_{i}%
\right\rangle _{\text{A}}=\left\langle s_{j}\right\rangle _{\text{A}}$ and
$\left\langle s_{i}s_{j}\right\rangle _{\text{A}}=\left\langle s_{i}%
\right\rangle _{\text{A}}\left\langle s_{j}\right\rangle _{\text{A}}$.
Therefore, we can write
\begin{align}
F_{\min}  &  =\epsilon_{\text{ND}}\left(  N-n-1\right)  \tanh\left(
\beta\epsilon_{\text{ND}}\right)  +\epsilon_{\text{D}}\left(  n\right)
\tanh\left(  \beta\epsilon_{\text{D}}\right)  -\frac{1}{\beta}\log\left[
2\cosh\left(  \beta\epsilon_{\text{ND}}\right)  \right]  ^{N-n-1}-\frac
{1}{\beta}\log\left[  2\cosh\left(  \beta\epsilon_{\text{D}}\right)  \right]
^{n-1}+\nonumber\\
& \nonumber\\
&  -J\sum_{i,j}^{N-n-1,n}\left\langle s_{i}\right\rangle _{\text{ND}%
}\left\langle s_{j}\right\rangle _{\text{ND}}-J\sum_{k,l}^{n,n-1}\left\langle
s_{k}\right\rangle _{\text{D}}\left\langle s_{l}\right\rangle _{\text{D}}-\mu
B^{\prime}\left[  \epsilon_{\text{ND}}\left(  N-n-1\right)  \tanh\left(
\beta\epsilon_{\text{ND}}\right)  +\epsilon_{\text{D}}\left(  n\right)
\tanh\left(  \beta\epsilon_{\text{D}}\right)  \right]  ~\text{.} \label{MM}%
\end{align}
Substituting Eqs. (\ref{S_ND}) and (\ref{S_D}) into Eq. (\ref{MM}) yields,
\begin{align}
F_{\min}  &  =\epsilon_{\text{ND}}\left(  N-n-1\right)  \tanh\left(
\beta\epsilon_{\text{ND}}\right)  +\epsilon_{\text{D}}\left(  n\right)
\tanh\left(  \beta\epsilon_{\text{D}}\right)  -\frac{1}{\beta}\log\left[
2\cosh\beta\epsilon_{\text{ND}}\right]  ^{N-n-1}+\nonumber\\
& \nonumber\\
&  -\frac{1}{\beta}\log\left[  2\cosh\beta\epsilon_{\text{D}}\right]
^{n-1}-\frac{1}{2}J\left(  N-n-1\right)  \left(  n\right)  \tanh^{2}\left(
\beta\epsilon_{\text{ND}}\right)  -\frac{1}{2}J\left(  n\right)  \left(
n-1\right)  \tanh^{2}\left(  \beta\epsilon_{\text{D}}\right)  +\nonumber\\
& \nonumber\\
&  -\mu B^{\prime}\left[  \epsilon_{\text{ND}}\left(  N-n-1\right)
\tanh\left(  \beta\epsilon_{\text{ND}}\right)  +\epsilon_{\text{D}}\left(
n\right)  \tanh\left(  \beta\epsilon_{\text{D}}\right)  \right]  \text{.}%
\end{align}
After collecting $\epsilon_{\text{ND}}$ and $\epsilon_{\text{D}}$ like terms,
we have
\begin{align}
F_{\min}  &  =\epsilon_{\text{ND}}\left(  N-n-1\right)  \tanh\left(
\beta\epsilon_{\text{ND}}\right)  -\frac{1}{\beta}\log\left[  2\cosh\left(
\beta\epsilon_{\text{ND}}\right)  \right]  ^{N-n-1}+\nonumber\\
& \nonumber\\
&  -\frac{1}{2}J\left(  N-n-1\right)  \left(  n\right)  \tanh^{2}\left(
\beta\epsilon_{\text{ND}}\right)  -\mu B^{\prime}\epsilon_{\text{ND}}\left(
N-n-1\right)  \tanh\left(  \beta\epsilon_{\text{ND}}\right)  +\nonumber\\
& \nonumber\\
&  +\epsilon_{\text{D}}\left(  n\right)  \tanh\left(  \beta\epsilon_{\text{D}%
}\right)  -\frac{1}{\beta}\log\left[  2\cosh\left(  \beta\epsilon_{\text{D}%
}\right)  \right]  ^{n-1}-\frac{1}{2}J\left(  n\right)  \left(  n-1\right)
\tanh^{2}\left(  \beta\epsilon_{\text{D}}\right)  -+\mu B^{\prime}%
\epsilon_{\text{D}}\left(  n\right)  \tanh\left(  \beta\epsilon_{\text{D}%
}\right)  ~\text{,}%
\end{align}
where the factor $n$ comes from the number of nearest neighbors, the factor
$N$ is from the total number of atoms and the $1/2$ is due to double counting.
We now choose the form that minimizes $F_{\min}$ with respect to
$\epsilon_{\text{ND}}$ and $\epsilon_{\text{D}}$. When minimizing with respect
to $\epsilon_{\text{ND}}$, we obtain
\begin{align}
\left.  \frac{\partial F_{\min}}{\partial\epsilon_{\text{ND}}}\right\vert
_{\epsilon_{\text{ND}}=~\epsilon_{\text{ND}\min}}  &  =0=\left(  N-n-1\right)
\tanh\left(  \beta\epsilon_{\text{ND}\min}\right)  +\left(  N-n-1\right)
\epsilon_{\text{ND}}\frac{\beta}{\cosh^{2}\left(  \beta\epsilon_{\text{ND}%
\min}\right)  }-\frac{N-n-1}{\beta}\frac{\beta\sinh\left(  \beta
\epsilon_{\text{ND}\min}\right)  }{\cosh\left(  \beta\epsilon_{\text{ND}\min
}\right)  }\nonumber\\
& \nonumber\\
&  -J\frac{1}{2}\left(  N-n-1\right)  \left(  n\right)  2\tanh\left(
\beta\epsilon_{\text{ND}\min}\right)  \frac{\beta}{\cosh^{2}\left(
\beta\epsilon_{\text{ND}\min}\right)  }-\mu B^{\prime}\left(  N-n-1\right)
\frac{\beta}{\cosh^{2}\left(  \beta\epsilon_{\text{ND}\min}\right)  }\text{,}%
\end{align}
that is, after some simple algebra,%
\begin{equation}
\epsilon_{\text{ND}\min}-\mu B^{\prime}=Jn\tanh\left(  \beta\epsilon
_{\text{ND}\min}\right)  \text{.} \label{fica1}%
\end{equation}
Similarly, when minimizing with respect to $\epsilon_{\text{D}}$, one gets
\begin{equation}
\epsilon_{\text{D}\min}-\mu B^{\prime}=J\left(  n-1\right)  \tanh\left(
\beta\epsilon_{\text{D}\min}\right)  \text{.} \label{fica2}%
\end{equation}

\begin{figure}[ptb]
\centering
\includegraphics[width=0.5\textwidth] {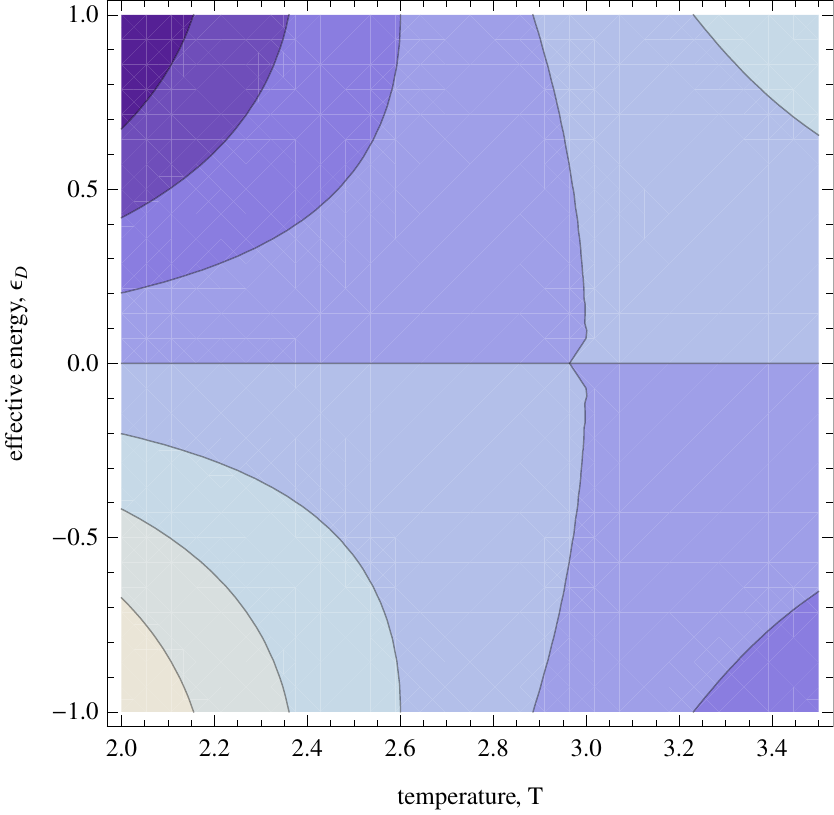}\caption{Contour plot exhibiting
the numerical solution of the effective energy of defective atoms
$\epsilon_{D}$ vs. temperature $T$. We set $k_{B}=1$ and assume $n=4$, $J=1$,
$\mu=1$, $B^{\prime}=1$ and the level curves $f\left(  \epsilon_{D}\text{,
}T\right)  =c$ with $c\in\left\{  -1.3\text{, }-1.2\text{, }-1.1\text{,
}-1\text{, }-0.9\right\}  $.}%
\label{fig5}%
\end{figure}

Eqs. (\ref{fica1}) and (\ref{fica2}) are final results. A contour plot of
$f\left(  \epsilon_{\text{D}}\text{, }T\right)  $\textbf{ }defined as,%
\begin{equation}
f\left(  \epsilon_{\text{D}}\text{, }T\right)  \overset{\text{def}}{=}%
\epsilon_{\text{D}}-\mu B^{\prime}-J\left(  n-1\right)  \tanh\left(
\frac{\epsilon_{\text{D}}}{k_{B}T}\right)  \text{,}%
\end{equation}
as a function of $\epsilon_{\text{D}}$ and $T$ appears in FIG. 5. These
equations can then be used as above to solve for the critical temperature, the
total magnetic moment and the magnetic susceptibility for this ferromagnetic
material as usual. The difference here is that we will have two of each. For
example, a critical temperature that applies to most of the atoms, and one
that is local to the defect. Therefore, the total magnetic moment will not
simply be a result of just one temperature but a combination of each magnetic
moment determined by these equations. Notice that Eq. (\ref{EA_G}) still holds
in general. In the two energy case, we have for one energy the number of
nearest neighbors, $n_{\text{G}}=n$ and for the second energy, $n_{\text{G}%
}=n-1$. As a final consideration, we point out that the outcomes of our
entropic analysis in Eqs. (\ref{fica1}) and (\ref{fica2}) together with the
numerical representation in FIG. 5 are consistent with the fact that
imperfections in the lattice structure can be the cause of structural
irregularities in the critical temperature $T_{c}$ \cite{jean00, kallin08}.
However, to obtain a deeper understanding of these aspects of ferromagnetic
materials from an entropic viewpoint, a more thorough analysis would be required.

Just as in the previous example, the\textbf{ }\textit{defect} could be
introduced here as an observable by means of a constraint like relation,%
\begin{equation}
\int dB\sum_{\left\{  s\neq s_{0}\right\}  }p\left\{  s,B\right\}  \equiv
p\left\{  s_{0}\right\}  =\delta_{s_{0}\text{, }s_{0}^{\prime}}\text{ ,}%
\end{equation}
where\textbf{ }$\delta_{s_{0}\text{, }s_{0}^{\prime}}$ is a Kronecker delta
function. This makes much more sense here as this information may be obtained
by common spectroscopic methods such as electron paramagnetic resonance,
photoluminescence or cathodoluminescence. In the particular case considered
here, we observed $s_{0}^{\prime}=0$.

\section{Conclusive Remarks}

In this article, we showed that not only can the MrE method be used for
updating probability distributions with both data and expectation values, but
for determining effective approximations of such probability distributions
equivalent to those provided by the Bogoliubov Variational Principle and/or
Mean Field Theory. Specifically, after describing the traditional Ising model
of a ferromagnetic material and the mean field approximation for a
multidimensional ferromagnet, we applied the MrE method to infer both
magnetization and energetic features of ferromagnets characterized by simple
lattice configurations. Our main results can be summarized as follows:

\begin{enumerate}
\item The standard mean field theory solution was recovered via MrE methods
for the ferromagnetic Ising model. These findings are reported in Eq.
(\ref{implicit}) and FIG. 2. The main advantage of our analysis is the fact
that we do not require the knowledge of the exact form of the effective energy
$\epsilon_{\text{A}}$ and we allow for departures from the centrality of the atom.

\item Moving past traditional methods and assumptions by using MrE, we
examined ferromagnets characterized by three state atoms. The outcomes of our
entropic analysis are reported in Eq. (\ref{cc}) and FIG. 3. Based on these
results, we concluded that our analysis appears to support the idea that the
introduction of additional states in the atomic ferromagnetic structure is
consistent with both space shifts and nonhomogeneities of the critical
temperature $T_{c}$ \cite{jean00, kallin08}.

\item A ferromagnetic material was considered\textbf{ }that had known defects,
which is beyond the scope of the mean field methodology. These findings are
presented in\ Eqs. (\ref{fica1}), (\ref{fica2}) and FIG. 5. In each case we
uncovered results independent of the explicit form of the effective energy in
the Hamiltonian. Relying on these results, we concluded that our research
appears to be consistent with the fact that imperfections in the lattice
structure can be the cause of structural irregularities in the critical
temperature $T_{c}$ \cite{jean00, kallin08}.

\item Although it may seem trivial to introduce the observational data through
an MrE constraint, as opposed to simply \textit{putting it in} as would be
conventionally done, it should be noted that these toy examples are mostly for
pedagogical purposes in order to demonstrate how the mechanism works. The
power of this methodology enters when one does not know the expectation value
of the Hamiltonian, but may know how it relates to some observable, as done in
\cite{giffin}. For instance, a more interesting example might be a case in
which the joint prior equals\textbf{ }$p_{\text{old}}\left\{  s,B\right\}
=p_{\text{old}}\left\{  s\right\}  p_{\text{old}}\{B~|~s\}$\textbf{
}where\textbf{ }$p_{\text{old}}\{B~|~s\}$\textbf{ }is the Bayesian likelihood
that represents a model that relates some observable\textbf{ }$B_{i}$\textbf{
}with\textbf{ }$s_{i}$\textbf{, }like\textbf{ }$p_{\text{old}}%
\{B~|~s\}\varpropto e^{-\beta\mathcal{H}_{ext}}$\textbf{ }with\textbf{
}$\mathcal{H}_{ext}$\textbf{ }described in Eq. (\ref{45}). Care must be taken
when examining this scenario and such considerations will be explored in later works.
\end{enumerate}

A function $f=f\left(  x\right)  $ can fail to be differentiable at a point
$x=x_{0}$ in three scenarios: it has a vertical tangent line at $x=x_{0}$, it
has a discontinuity at $x=x_{0}$, it exhibits a kink (or, corner) at $x=x_{0}%
$. Interestingly, we point out that Figs. $2$, $3$, and $5$ show a vertical
tangent, a jump discontinuity, and a kink, respectively. Specifically, the
arbitrary function $f$ is represented by the mean magnetization of the
ferromagnetic material in FIG. $2$, by the effective energy of an atom in FIG.
$3$, and finally, by the effective energy of a defective atom in FIG. $5$.
Although numerical methods are often ineffective near points of discontinuity,
our qualitative entropic analysis seems to suggest that departures from the
standard Ising model (with a vertical tangent type of singularity) can lead to
alternative types of non-differentiable singularities.

While the findings uncovered in this manuscript are not conclusive (about the
nature of phase transitions in the presence of imperfections) by any means,
this work seems to suggest that using the MrE method can be useful in
inferring both thermal and electromagnetic properties of defective
ferromagnetic materials in presence of available data from the substance being
considered. Of course, a more detailed investigation would be required to
deepen our understanding of these important phenomena in fundamental and
applied science. Indeed, it is our intention to extend our study to higher
order spin configurations \cite{Stanley} (for example, XY spin chains, XX spin
chains, Heisenberg models, or, more generally, Potts models \cite{Potts, wu}).
Finally, we point out that MrE methods employed in this article can be readily
applied, in principle, to the investigation of more general dislocation,
disclination, and dispiration defects in solid state physics \cite{ali07}. In
particular, it would be very interesting to infer physical properties of
materials exhibiting a non-uniform distribution of defects of different types.
We leave these investigations to future efforts.

\begin{acknowledgments}
A. Giffin would like to acknowledge valuable discussions with Ariel Caticha
regarding some of the material presented in this manuscript. Finally,
constructive criticisms from an anonymous referee leading to an improved
version of this manuscript are sincerely acknowledged by the authors.
\end{acknowledgments}


\begin{thebibliography}{99}                                                                                               %


\bibitem {Jaynes}E. T. Jaynes, \emph{Information theory and statistical
mechanics}, Phys. Rev. \textbf{106}, 620 (1957).

\bibitem {Jaynes2}E. T. Jaynes, \emph{Information theory and statistical
mechanics II}, Phys. Rev. \textbf{108}, 171 (1957).

\bibitem {Shannon1948}C. E. Shannon, \emph{A mathematical theory of
communication}, \emph{Bell System Technical Journal} \textbf{27}, 379 (1948).

\bibitem {CatichaGiffin}A. Caticha and A. Giffin,\emph{ Updating
probabilities}, AIP Conf. Proc. \textbf{872}, 31 (2006).

\bibitem {Tseng}C.Y. Tseng, \emph{The Maximum Entropy Method in Statistical
Physics: an Alternative Approach to the Theory of Simple Fluids}, Ph. D.
Thesis, SUNY at Albany, Albany-New York, USA (2004).

\bibitem {Reif}F. Reif, \emph{Fundamentals of Statistical and Thermal
Physics}, McGraw-Hill (1965).

\bibitem {lieb64}T. D. Schultz, D.\ C. Mattis, and E. H. Lieb,
\emph{Two-dimensional Ising model as a soluble problem of many fermions}, Rev.
Mod. Phys. \textbf{36}, 856 (1964).

\bibitem {brush67}S. G. Brush, \emph{History of the Lenz-Ising model}, Rev.
Mod. Phys. \textbf{39}, 883 (1967).

\bibitem {cipra00}B. A. Cipra, \emph{The Ising model is NP-complete},\ SIAM
News \textbf{33}, 6 (2000).

\bibitem {huang}K. Huang, \emph{Statistical Mechanics}, John Wiley \& Sons,
Inc. (1963).

\bibitem {mermin}N. W. Ashcroft and N. D. Mermin, \emph{Solid State Physics},
Thomson Learning, Inc. (1976).

\bibitem {Schroeder}D. V. Schroeder, \emph{An Introduction to Thermal
Physics}, Addison Wesley Longman (2000).

\bibitem {carlo1}C. Cafaro and S. A. Ali, \emph{Jacobi fields on statistical
manifolds of negative curvature}, Physica \textbf{D234}, 70 (2007).

\bibitem {carlo2}C. Cafaro, \emph{Information geometry, inference methods and
chaotic energy levels statistics}, Mod. Phys. Lett. \textbf{B22}, 1879 (2008).

\bibitem {carlo3}C. Cafaro and S. A. Ali, \emph{Can chaotic quantum energy
levels statistics be characterized using information geometry and inference
methods?}, Physica \textbf{A387}, 6876 (2008).

\bibitem {carlo4}C. Cafaro and S. A. Ali, \emph{Geometrodynamics of
information on curved statistical manifolds and its applications to chaos},
EJTP \textbf{5}, 139 (2008).

\bibitem {carlo5}C. Cafaro, A. Giffin, S. A. Ali, and D.-H. Kim,
\emph{Reexamination of an information geometric construction of entropic
indicators of complexity}, Appl. Math. Comput. \textbf{217}, 2944 (2010).

\bibitem {carlo6}C. Cafaro and S. Mancini, \emph{Quantifying the complexity of
geodesic paths on curved statistical manifolds through information geometric
entropies and Jacobi fields}, Physica \textbf{D240}, 607 (2011).

\bibitem {carlo7}D.-H. Kim, S. A. Ali, C. Cafaro\textbf{,} and S. Mancini,
\emph{Information geometric modeling of scattering induced quantum
entanglement}, Phys. Lett. \textbf{A375}, 2868 (2011).

\bibitem {carlo8}C. Cafaro, A. Giffin, C. Lupo, and S. Mancini,
\emph{Softening the complexity of entropic motion on curved statistical
manifolds}, Open Systems \& Information Dynamics \textbf{19}, 1250001 (2012).

\bibitem {carlo9}D. Felice, C. Cafaro, and S. Mancini,\emph{\ Information
geometric complexity of a trivariate Gaussian statistical model}, Entropy
\textbf{16}, 2944 (2014).

\bibitem {sadoc06}J. F. Sadoc and R. Mosseri, \emph{Geometrical Frustration},
Cambridge University Press (2006).

\bibitem {giffin-caticha}A. Giffin and A. Caticha, \emph{Updating
probabilities with data and moments}, AIP Conf. Proc. \textbf{954}, 74 (2007).

\bibitem {giffin}A. Giffin, \emph{From physics to economics: An econometric
example using maximum relative entropy}, Physica \textbf{A388}, 1610 (2009).

\bibitem {tseng2008}C.-Y. Tseng and A. Caticha, \emph{Using relative entropy
to find optimal approximations: an application to simple fluids}, Physica
\textbf{A387}, 6759 (2008).

\bibitem {kuzemsky}A. L. Kuzemsky, \emph{Variational principle of Bogoliubov
and generalized mean fields in many-particle interacting systems}, Int. J.
Mod. Phys. \textbf{B29}, 1530010 (2015).

\bibitem {onsager}L. Onsager, \emph{Crystal statistics. I. A two-dimensional
model with an order-disorder transition}, Phys. Rev. \textbf{65}, 117 (1944).

\bibitem {yang}C. N. Yang, \emph{The spontaneous magnetization of a
two-dimensional Ising model}, Phys. Rev. \textbf{85}, 808 (1952).

\bibitem {ising}E. Ising, \emph{Beitrag zur theorie des ferromagnetismus},
Zeitschrift fur Physik\textbf{ 31}, 253 (1925).

\bibitem {weiss}P. Weiss, \emph{L'hypothese du champ moleculaire et la
propriete ferromagnetique}, J. Phys. Theor. Appl. \textbf{6}, 661 (1907).

\bibitem {mac}D. J. C. MacKay,\emph{ Information Theory, Inference, and
Learning Algorithms}, Cambridge University Press (2003).

\bibitem {stanleybook}H. E. Stanley, \emph{Introduction to phase transitions
and critical phenomena}, Oxford University Press (1971).

\bibitem {Caticha}A. Caticha, \emph{Entropic Inference and the Foundations of
Physics}, USP Press, Sao Paulo, Brazil (2012).

\bibitem {jean00}F. Jean, G. Collin, M. Andrieux, N. Blanchard, J.-F. Marucco,
\emph{Oxygen nonstoichiometry, point defects and critical temperature in
superconducting oxide Bi}$_{2}$\emph{Sr}$_{2}$\emph{CaCu}$_{2}$\emph{O}%
$_{8+\Delta}$, Physica \textbf{C339}, 269 (2000).

\bibitem {kallin08}D. Davidovich, A. J. Berlinsky, and C. Kallin,
\emph{Superfluid density near the critical temperature in the presence of
random planar defects}, Phys. Rev. \textbf{B78}, 214508 (2008).

\bibitem {ashkin1943}J. Ashkin and E. Teller, \emph{Statistics of
two-dimensional lattices with four components}, Phys. Rev. \textbf{64}, 178 (1943).

\bibitem {Stanley}H. E. Stanley, \emph{Dependence of critical properties upon
dimensionality of spins}, Phys. Rev. Lett. \textbf{20}, 589 (1968).

\bibitem {Potts}R. B. Potts, \emph{Some Generalized Order-Disorder
Transformations}, Mathematical Proceedings \textbf{48}, 106 (1952).

\bibitem {wu}F. Y. Wu, \emph{The Potts model}, Rev. Mod. Phys.\textbf{ 54},
235 (1982).

\bibitem {ali07}S. A. Ali, C. Cafaro, S. Capozziello, and Ch. Corda, \emph{A
bound quantum particle in a Riemann-Cartan space with topological defects and
planar potential}, Phys. Lett. \textbf{A366}, 315 (2007).
\end{thebibliography}
\end{document}